\documentclass[preprint]{JHEP3} % 10pt is ignored!

\JHEPspecialurl{http://jhep.sissa.it/JOURNAL/JHEP3.tar.gz}

\usepackage{epsfig,multicol,bbm}
\usepackage{longtable}
\usepackage{amsthm,latexsym,multibox,amssymb,amsfonts,array}
\usepackage[centertags,intlimits]{amsmath}
\usepackage{graphicx}
\usepackage{epstopdf}
\newcommand\fverb{\setbox\pippobox=\hbox\bgroup\verb}
\newcommand\fverbdo{\egroup\medskip\noindent%
            \fbox{\unhbox\pippobox}\ }
\newcommand\fverbit{\egroup\item[\fbox{\unhbox\pippobox}]}
\newbox\pippobox

\newcommand{\al}{\alpha}

\newcommand{\beq}{\begin{equation}}
\newcommand{\eeq}{\end{equation}}
\newcommand{\beqa}{\begin{eqnarray}}
\newcommand{\eeqa}{\end{eqnarray}}
\newcommand{\eqnlab}[1]{\label{eqn:#1}}
\newcommand{\Eqnref}[1]{Eq.~(\ref{eqn:#1})}
\newcommand{\f}{\frac}
\newcommand{\mc}{\mathcal}

\newcommand{\hs}{\hspace{0.1 cm}}

\newcommand{\be}{\beta}
\newcommand{\ga}{\gamma}
\newcommand{\de}{\delta}

\newcommand{\nn}{\nonumber}
\newcommand{\ph}{\phantom}
\newcommand{\mf}{\mathfrak}

\newcommand{\mbb}{\mathbb}
\newcommand{\vg}{\vec{g}}
\newcommand{\vp}{\vec{\phi}}
\newcommand{\gp}{\vec{g}\cdot\vec{\phi}}
\newcommand{\vf}{\vec{f}}
\newcommand{\fip}{\vec{f}_i\cdot\vec{\phi}}
\newcommand{\ve}{\vec{e}}
\newcommand{\vo}{\vec{\omega}}
\newcommand{\val}{\vec{\al}}
\newcommand{\vL}{\vec{\Lambda}}
\newcommand{\vl}{\vec{\lambda}}
\newcommand{\noi}{\noindent}
\newcommand{\vscalar}[2]{\vec{#1}\cdot\vec{#2}}
\newcommand{\gphi}{\vec{g}\cdot\vec{\phi}}
\newcommand{\pa}{\partial}
\newcommand{\fphi}[1]{\vec{f}_{#1}\cdot\vec{\phi}}
\newcommand{\Pc}{\mathcal{P}}
\newcommand{\vb}[2]{e_{#1}^{\phantom{#1}{#2}}}
\newcommand{\hvb}[2]{\hat{e}_{#1}^{\phantom{#1}{#2}}}
\newcommand{\tvb}[2]{{\tilde{e}}_{#1}^{\phantom{#1}{#2}}}
\newcommand{\ivb}[2]{u_{#1}^{\phantom{#1}{#2}}}
\newcommand{\tvbint}{\tilde{e}_{\mathrm{int}}}

\newcommand{\tF}{\tilde{F}}
\newcommand{\tP}{\tilde{P}}

\newcommand{\tH}{\tilde{H}}
\newcommand{\tr}{\mathrm{tr}}
\newcommand{\trtP}{\mathrm{tr}\tilde{P}}

\newcommand{\tFPPFa}{\tF^a_{\phantom{a}\gamma\delta} \tP_{\epsilon ab}
\tP^{\epsilon bc}\tF_c^{\phantom{c}\gamma\delta}}
\newcommand{\tFPPFb}{\tF^c_{\phantom{c}\beta\gamma}
\tP^\gamma_{\phantom{\gamma} ce}\tP_\delta^{\phantom{\delta}
ed}\tF_d^{\phantom{d}\beta\delta}}
\newcommand{\tFPPFc}{\tF^c_{\phantom{c}\beta\gamma} \tP_{\delta ce}
\tP^{\gamma ed}\tF_d^{\phantom{d}\beta\delta}}

\newcommand{\tFDPF}{\tF_{c\alpha\beta}(D_\gamma\tP^{\gamma
cd})\tF_d^{\phantom{d}\alpha\beta}}

\newcommand{\tFFFFa}{\tF_{c\alpha\beta}\tF_d^{\phantom{d}\alpha\beta}
\tF^c_{\phantom{c}\gamma\delta}\tF^{d\gamma\delta}}
\newcommand{\tFFFFb}{\tF_{c\gamma\alpha}
\tF_{\phantom{c\gamma}\beta}^{c\gamma}
\tF^{\phantom{d\delta}\alpha}_{d\delta}\tF^{d\delta\beta}}
\newcommand{\tFFFFc}{\tF_{c\alpha\beta}
\tF^c_{\phantom{c}\gamma\delta} \tF_d^{\phantom{d}\alpha\gamma}
\tF^{d\beta\delta}}

\newcommand{\tFPFa}{\tF_{c\beta\gamma}\tP^{\alpha
cd}\tF_d^{\phantom{d}\beta\gamma}}
\newcommand{\tFPFb}{\tF_{c\beta\gamma}\tP^{\gamma
cd}\tF_d^{\phantom{d}\beta\alpha}}

\newcommand{\tHHHH}{\tH_{a\gamma}\tH_b^{\phantom{b}\gamma}
\tH^a_{\phantom{a}\delta}\tH^{b\delta}}
\newcommand{\tHPPHa}{\tH^{c\alpha}\tP_{\alpha cd} \tP^{\beta
de}\tH_{e \beta}}
\newcommand{\tHPPHb}{\tH^{c\alpha}\tP_{\beta cd} \tP^{\beta
de}\tH_{e \alpha}}
\newcommand{\tHPHa}{\tH_{c\alpha}\tP^{\alpha cd}
\tH_d^{\phantom{d}\beta}}
\newcommand{\tHPHb}{\tH_{c\alpha}\tP^{\beta cd}
\tH_d^{\phantom{d}\alpha}}

\newcommand{\EOMtF}{D_\alpha \tF_{c\delta}^{\phantom{c\delta}\alpha}
+ \tP_{\alpha c}^{\phantom{\alpha
c}e}\tF_{e\delta}^{\phantom{e\delta}\alpha}}

\newcommand{\EOMtPa}{D_\alpha \tP^\alpha_{\phantom{\alpha}cd} -
\frac{1}{4}\tF_{c\alpha\beta}\tF_d^{\phantom{d}\alpha\beta} -
\frac{1}{(D-2)}\delta_{cd} \tr(D_\alpha \tP^\alpha)}

\newcommand{\EOMtrtP}{\tr(D_\alpha \tP^\alpha) -
\frac{(D-2)}{4(D-n-2)}\tF^2}

\newcommand{\EOMphi}{\square\varphi + \frac{1}{4(D-n-2)}\tF^2}

\preprint{G\"oteborg preprint \\ ULB-TH/07-30}

\title{U-Duality and the Compactified Gauss-Bonnet Term}

\author{Ling BAO, Johan BIELECKI, Martin CEDERWALL and Bengt E. W. NILSSON \\

Fundamental Physics \\
Chalmers University of Technology,\\
SE 412 96 G\"oteborg, Sweden\\

E-mail: \email{ling.bao@chalmers.se},
\email{bieljoha@student.chalmers.se},
\email{martin.cederwall@chalmers.se}, \email{tfebn@fy.chalmers.se}}

\author{Daniel PERSSON\\

Physique Th\'{e}orique et Math\'{e}matique, \\ Universit\'{e} Libre
de Bruxelles \&
International Solvay Institutes, \\
ULB-Campus Plaine C.P.231, B-1050 Bruxelles,
Belgium.\\

E-mail: \email{dpersson@ulb.ac.be}}

\abstract{We present the complete toroidal compactification of the
Gauss-Bonnet Lagrangian from $D$ dimensions to $D-n$ dimensions. Our
goal is to investigate the resulting action from the point of view
of the ``U-duality'' symmetry $SL(n+1, \mbb{R})$ which is present in
the tree-level Lagrangian when $D-n=3$. The analysis builds upon and
extends the investigation of the paper [arXiv:0706.1183], by
computing in detail the full structure of the compactified
Gauss-Bonnet term, including the contribution from the  dilaton
exponents. We analyze these exponents using the representation
theory of the Lie algebra $\mf{sl}(n+1, \mbb{R})$ and determine
which representation seems to be the relevant one for quadratic
curvature corrections. By interpreting the result of the
compactification as a leading term in a large volume expansion of an
$SL(n+1, \mbb{Z})$-invariant action, we conclude that the overall
exponential dilaton factor should not be included in the
representation structure. As a consequence, all dilaton exponents
correspond to weights of $\mf{sl}(n+1, \mbb{R})$, which,
nevertheless, remain on the positive side of the root lattice.}

\begin{document}

%\maketitle  IS IGNORED %%%%%%%%%%%
\section{Introduction and Summary}

Dimensional reduction of supergravity theories is an efficient
method of revealing symmetry structures which are ``hidden'' when
the theories are formulated in maximal dimension. The first
discovery of such a hidden symmetry was the so-called Ehlers
symmetry of pure four-dimensional gravity compactified on a circle
to three dimensions \cite{Ehlers}. The global symmetry $GL(1,
\mbb{R})=\mbb{R}$, corresponding to rescaling of the $S^1$, is in
this case extended through dualisation of the Kaluza-Klein vector
into a new scalar, revealing that the full global symmetry of the
Lagrangian is, in fact, described by the group $SL(2, \mbb{R})$. The
scalars in the theory parametrise the coset space $SL(2,
\mbb{R})/SO(2)$, where $SO(2)$ is the maximal compact subgroup of
$SL(2, \mbb{R})$, playing the role of a local gauge symmetry. More
generally, upon toroidal compactification of lowest order pure
gravity in $D$ spacetime dimensions on an $n$-torus, $T^n$, to three
dimensions, the scalars parametrise the coset space $SL(n+1,
\mbb{R})/SO(n+1)$. The enhancement from $GL(n, \mbb{R})$ to $SL(n+1,
\mbb{R})$ is again due to the fact that in three dimensions all
Kaluza-Klein vectors can be dualised to scalars.

Similar phenomena occur also for coupled gravity-dilaton-$p$-form
theories, such as the bosonic sectors of the low-energy effective
actions of string and M-theory. The most thoroughly investigated
case is the toroidal compactification of eleven-dimensional
supergravity on $T^n$ to $d=11-n$ dimensions, for which the scalar
sector parametrises the coset space
$\mc{E}_{n(n)}/\mc{K}(\mc{E}_{n(n)})$, with $\mc{K}(\mc{E}_{n(n)})$
being the (locally realized) maximal compact subgroup of
$\mc{E}_{n(n)}$ \cite{CremmerJulia}. In particular, for reduction to
three dimensions the global symmetry group is the split real form
$\mc{E}_{8(8)}$, with maximal compact subgroup $Spin(16)/\mbb{Z}_2$.
The global symmetry group $\mc{E}_{8(8)}$ is the U-duality group,
which, from a string theory perspective, combines the
non-perturbative S-duality group $SL(2, \mbb{R})$ of type IIB
supergravity with the perturbative T-duality group $SO(7, 7)$
\cite{UDualityReview}.

These symmetries are present in the classical (tree-level)
Lagrangian, but it is known from string theory that they must be
broken by quantum effects. It has been conjectured that if
$\mc{U}_d$ is the continuous symmetry group appearing upon
compactification from $D$ to $d=D-n$ dimensions, then a discrete
subgroup $\mc{U}_d(\mbb{Z})\subset \mc{U}_d$ lifts to a symmetry of
the full quantum theory \cite{HullTownsend}.\footnote{Strictly
speaking, the name U-duality is reserved for the chain of
exceptional discrete groups $\mc{E}_{n(n)}(\mbb{Z})$, related to the
toroidal compactification of M-theory (see \cite{UDualityReview} for
a review). However, for convenience, we shall in this paper adopt a
slight abuse of terminology and refer to any enhanced symmetry group
$\mc{U}_d(\mbb{Z})$ as a ``U-duality'' group. This then applies, for
example, to the mapping class group $SL(n+1, \mbb{Z})$ of the
internal torus in the reduction of pure gravity to three dimensions,
and to the T-duality group $SO(n, n, \mbb{Z})$ appearing in the
reduction of the coupled gravity-2-form system. Moreover, we shall
refer to the continuous versions of these groups,
$\mc{U}_d=\mc{U}_d(\mbb{R})$, as ``classical U-duality groups''.}
The physical degrees of freedom of the scalar sector then
parametrise the coset space
$\mc{U}_d(\mbb{Z})\verb|\|\mc{U}_d\verb|/|\mc{K}(\mc{U}_d)$.

\subsection{Non-Perturbative Completion and Automorphic Forms}

Recently, several authors
\cite{LambertWest1,LambertWest2,Aspects,ClaudiaAxel,PiolineYann}
have initiated an investigation aimed at answering the question of
whether or not the U-duality group $\mc{U}_3$ in three dimensions is
preserved also if the tree-level Lagrangian is supplemented by
higher order curvature corrections. The consensus has been that
toroidal compactifications of quadratic and higher order corrections
give rise to terms which are not $\mc{U}_3$-invariant.\footnote{One
exception being ref. \cite{ClaudiaAxel} in which the authors
considered quadratic curvature corrections to pure gravity in four
dimensions. In that special case, the most general correction can be
related, through suitable field redefinitions, to the Gauss-Bonnet
term which is topological in four dimensions and does not contribute
to the dynamics. Hence, the $SL(2, \mbb{R})$-symmetry of the
compactified Lagrangian is trivially preserved.}

A nice example of a fairly well understood realisation of these
mechanisms is the breaking of the classical $SL(2, \mbb{R})$
symmetry of the type IIB supergravity effective action down to the
quantum S-duality group $SL(2, \mbb{Z})$ of the full type IIB string
theory \cite{GreenGutperle}. The next to leading order
$\al^{\prime}$-corrections to the effective action are octic in
derivatives of the metric, i.e., fourth order in powers of the
Riemann tensor, and receives perturbative contributions only from
tree-level and one-loop in the string genus expansion. However, this
gives a scalar coefficient in front of the $\mc{R}^4$-terms in the
effective action which is not $SL(2, \mbb{Z})$-invariant. This
problem is resolved by noting that there are additional
non-perturbative contributions to the octic derivative terms arising
from $D$-instantons ($D(-1)$-branes) \cite{GreenGutperle}. This
contribution can be seen as a ``completion'' of the coefficient to
an $SL(2, \mbb{Z})$-invariant scalar function which is identified
with a certain automorphic function, known as a non-holomorphic
Eisenstein series. A weak-coupling (large volume) expansion of this
function reproduces the perturbative tree-level and one-loop
coefficients at lowest order.

In the scenario described above the completion to a U-duality
invariant expression was achieved through the use of a scalar
automorphic form, i.e., an automorphic function, which is completely
$SL(2, \mbb{Z})$-invariant. More generally, one might find terms in
the effective action whose non-perturbative completion requires
automorphic forms transforming under the maximal compact subgroup
$\mc{K}(\mc{U}_3)$. For example, this was found to be the case in
\cite{GreenGutperleKwon}, where interaction terms of sixteen
fermions were analyzed. These terms transform under the maximal
compact subgroup $U(1)\subset SL(2, \mbb{R})$ and so the U-duality
invariant completion requires in this case an automorphic form which
transform with a $U(1)$ weight that compensates for the
transformation of the fermionic term, and thus renders the effective
action invariant.

The need for automorphic forms which transform under the maximal
compact subgroup $\mc{K}(\mc{U}_3)$ was also emphasized in
\cite{LambertWest2}, based on the observation that the dilaton
exponents in compactified higher curvature corrections correspond to
weights of the global symmetry group $\mc{U}_3$, implying that these
terms transform non-trivially in some representation of
$\mc{K}(\mc{U}_3)$. An explicit realisation of these arguments was
found in \cite{PiolineYann} for the case of compactification on
$S^1$ of the four-dimensional coupled Einstein-Liouville system,
supplemented by a four-derivative curvature correction. The
resulting effective action was shown to explicitly break the Ehlers
$SL(2, \mbb{R})$-symmetry; however, an $SL(2,
\mbb{Z})_{\text{global}}\times U(1)_{\text{local}}$-invariant
effective action was obtained by ``lifting'' the scalar coefficients
to automorphic forms transforming with compensating $U(1)$ weights.
The non-perturbative completion implied by this lifting is in this
case attributed to gravitational Taub-NUT instantons
\cite{PiolineYann}.

Similar conclusions were drawn in \cite{Aspects}, in which
compactifications of derivative corrections of second, third and
fourth powers of the Riemann tensor were analyzed. Again, it was
concluded that the $\mc{U}_3$-symmetry is explicitly broken by the
correction terms. It was argued, in accordance with the type IIB
analysis discussed above, that the result  of the compactification
-- being inherently perturbative in nature -- should be considered
as the large volume expansion of a $\mc{U}_3(\mbb{Z})$-invariant
effective action. It was shown on general grounds that any term
resulting from such a compactification can always be lifted to a
U-duality invariant expression through the use of automorphic forms
transforming in some representation of $\mc{K}(\mc{U}_3)$.

In this paper we extend some aspects of the analysis of
\cite{Aspects}. In \cite{Aspects} only parts of the compactification
of the Riemann tensor squared, $\hat{R}_{ABCD}\hat{R}^{ABCD}$, were
presented. The terms which were analyzed were sufficient to show
that the continuous symmetry was broken, and to argue for the
necessity of introducing transforming automorphic forms to restore
the U-duality symmetry $\mc{U}_3(\mbb{Z})$. Moreover, the overall
volume factor of the internal torus was  neglected in the analysis.

We restrict our study to corrections quadratic in the Riemann tensor
in order for a complete compactification to be a feasible task. More
precisely, we shall focus on a four-derivative correction to the
Einstein-Hilbert action in the form of the Gauss-Bonnet term
$\hat{R}_{ABCD}\hat{R}^{ABCD}-4\hat{R}_{AB}\hat{R}^{AB}+\hat{R}^2$.
Modulo field equations, this is the only independent invariant
quadratic in the Riemann tensor. We extend the investigations of
\cite{Aspects} by giving the complete compactification on $T^n$ of
the Gauss-Bonnet term from $D$ dimensions to $D-n$ dimensions. In
the special case of compactifications to $D-n=3$ dimensions the
resulting expression simplifies, making it amenable for a more
careful analysis. In particular, one of the main points of this
paper is to study the full structure of the dilaton exponents, with
the purpose of determining the $\mf{sl}(n+1,\mbb{R})$-representation structure associated with quadratic
curvature corrections. In contrast to the general arguments of
\cite{LambertWest1}Ê we have here access to a complete expression
after compactification, thus allowing us to perform an exhaustive
analysis of the weight structure associated with all terms in the
Lagrangian.

We note that effects of adding Gauss-Bonnet correction terms have recently been discussed in the contexts of black hole entropy (see \cite{Sen} for a recent review and further references) and brane world scenarios (see, e.g., \cite{Charmousis}).

\subsection{A Puzzle and a Possible Resolution}

The research programme outlined above was initially inspired by
recent results regarding the question of how curvature corrections
in string and M-theory, analyzed close to a spacelike singularity
(the ``BKL-limit''), fit into the representation structure of the
hyperbolic Kac-Moody algebra $E_{10(10)}=\text{Lie}\
\mc{E}_{10(10)}$ \cite{DamourNicolai,DHHKN}. These authors found
that generically such curvature corrections are associated with
exponents which reside on the negative side of the root lattice of
the algebra, indicating that correction terms fall into
infinite-dimensional (non-integrable) lowest-weight representations
of $E_{10(10)}$.\footnote{The root lattice of $E_{10(10)}$ is
self-dual, implying that the root lattice and the weight lattice
coincide. The same is true for $E_{8(8)}$.} Moreover, it was shown
that curvature corrections to eleven-dimensional supergravity match
with the root lattice of $E_{10(10)}$ only for the special powers
$3k+1,\ k=1, 2, 3, \dots, $ of the Riemann tensor. This is in perfect
agreement with explicit loop calculations, which reveal that the
only correction terms with non-zero coefficients are $\mc{R}^4,
\mc{R}^{7},\dots$, etc. \cite{GreenVanhove}. However, when reducing
to ten-dimensions and repeating the analysis for type IIA and type
IIB supergravity, the restriction on the curvature terms -- obtained
by requiring compatibility with the $E_{10(10)}$-root lattice -- no
longer match with known results from string calculations
\cite{DHHKN}. For example, the $E_{10(10)}$ analysis for type IIA
predicts a correction term of order $\mc{R}^3$, which is known to be
forbidden by supersymmetry. This implies that -- even though correct
for eleven-dimensional supergravity -- the compatibility between
higher derivative corrections and the root lattice of $E_{10(10)}$
is clearly not well-understood, and requires refinement.

These results are puzzling also in other respects, most notably
because the weights that arise from curvature corrections are
\emph{negative} weights of $E_{10(10)}$; with the leading order term
in a BKL-like expansion of the $\mc{R}^4$-terms being the lowest
weight of the representation, and, in fact, corresponds to the
negative of a dominant integral weight. This implies that the
representation builds upwards and outwards from the interior of the
negative fundamental Weyl chamber, rendering the representation
non-integrable. From the point of view of the nonlinear sigma model
for $\mc{E}_{10(10)}/\mc{K}(\mc{E}_{10(10)})$ this result is also
strange, because the correspondence with the tree-level Lagrangian
in the BKL-limit requires the use of the Borel gauge, for which no
negative weights appear in the Lagrangian \cite{DHN} (see
\cite{DHNReview,LivingReview} for reviews). The reason for these
puzzling results is essentially due to the ``lapse-function'' $N$,
representing the reparametrisation invariance in the timelike
direction. At tree-level the powers of the lapse-function arising
from the measure and from the Ricci scalar cancel, and the remaining
exponents correspond to positive roots of $E_{10(10)}$. On the other
hand, for terms of higher order in the Riemann tensor there are also
higher powers of the lapse-function which ``pushes'' the exponents
to the negative side of the root system.

From a different point of view, similar features have appeared in
the analysis of \cite{LambertWest1}. These authors investigated the
general structure of the dilaton exponents upon compactifications on
$T^8$ of quartic curvature corrections to eleven-dimensional
supergravity, emphasizing the importance of including the overall
``volume factor'', which parametrises the volume of the internal
torus. Of course, in this case it is the Lie algebra
$E_{8(8)}=\text{Lie}\ \mc{E}_{8(8)}$ which is the relevant one,
rather than $E_{10(10)}$. However, the inclusion of the volume
factor into the dilaton exponents when investigating the weight
structure has precisely the same effect as the lapse-function had in
the $E_{10(10)}$-case above, namely to push the exponents from the
positive root lattice of $E_{8(8)}$ down to the negative root
lattice, thus giving rise to negative weights of $E_{8(8)}$.

These results imply that one might use the simpler approach of
compactification of curvature corrections to three dimensions in
order to develop some intuition regarding the more difficult case of
implementing the full $\mc{E}_{10(10)}$-symmetry in M-theory. Based
on these considerations -- and the results obtained in the present
paper concerning the representation structure of the compactified
Gauss-Bonnet term -- we shall in fact argue that the overall volume
factor should \emph{not} be included in the analysis of the
representation structure. This interpretation draws from the idea
that the result of the compactification should be seen as the lowest
order term in a large volume expansion of a manifestly U-duality
invariant action. From this point of view the volume factor is then
associated to the first term in an expansion of an automorphic form
of $\mc{U}_3(\mbb{Z})$, transforming in some representation of the
maximal compact subgroup $\mc{K}(\mc{U}_3)$. Moreover, with this
interpretation, the dilaton exponents of the compactified quadratic
corrections exhibit a more natural structure in terms of
representations of $\mc{U}_3$. It is our hope that these results can
also be applied to the question of how higher derivative corrections
to eleven-dimensional supergravity fit into $E_{10(10)}$.

\subsection{Organisation of the Paper}

Our paper is organized as follows. In Section
\ref{section:Compactification} we present the result of the
compactification of the Gauss-Bonnet term on $T^n$ from $D$
dimensions to $D-n=3$ dimensions. The completely general action
representing the compactification to arbitrary dimensions is given
in Appendix \ref{appendix}. The result in three dimensions is given
in Section \ref{section:Compactification} after dualisation of all
Kaluza-Klein vectors into scalars, which is the case of most
interest from the U-duality point of view. We then proceed in
Section \ref{section:AlgebraicStructure} with the analysis of the
compactified Lagrangian. We analyze in detail the dilaton exponents
in terms of the representation theory of $\mf{sl}(n+1, \mbb{R})$,
which is the enhanced symmetry group of the compactified tree-level
Lagrangian. Finally, in Section \ref{section:resolution} we suggest
a possible non-perturbative completion of the compactified
Lagrangian into a manifestly U-duality invariant expression. We
explain how this completion requires the lifting of the coefficients
in the Lagrangian into automorphic forms transforming non-trivially
under the maximal compact subgroup $\mc{K}(\mc{U}_3)\subset
\mc{U}_3$. We interpret our results and provide a comparison with
the existing literature. All calculational details are displayed in
Appendix \ref{appendix}.

%%%%%%%%%%%%%%%%%%%%%%%%%%%%%%%%%%%%%%%%%%%%%%%%
\section{Compactification of the Gauss-Bonnet Term}

\label{section:Compactification}

In this section we outline the derivation of the toroidal
compactification of the Gauss-Bonnet term from $D$ dimensions to
$D-n$ dimensions. In \Eqnref{GBtilde} of Appendix \ref{appendix} we
give the full result for the compactification to arbitrary
dimensions. Here we focus on the special case of $D-n=3$, which is
the most relevant case for the questions we pursue in this paper.

\subsection{The General Procedure}

The Gauss-Bonnet Lagrangian density is quadratic in the Riemann
tensor and takes the explicit form
\beq
\mc{L}_{\mathrm{GB}}=\hat{e}\big[\hat{R}_{ABCD}\hat{R}^{ABCD} -
4\hat{R}_{AB}\hat{R}^{AB}+\hat{R}^2\big].
\eqnlab{GaussBonnet}
\eeq
The compactification of the $D$-dimensional Riemann tensor
${\hat{R}^{A}}_{BCD}$ on an $n$-torus, $T^n$, is done in three
steps: first we perform a Weyl-rescaling of the total vielbein,
followed by a splitting of the external and internal indices, and
finally we define the parametrisation of the internal vielbein. In
the following we shall always assume that the torsion vanishes.

\subsubsection*{Conventions and Reduction Ansatz}

Our index conventions are as follows. $M,N,\dots $ denote $D$
dimensional curved indices, and $A, B, \dots$ denote $D$ dimensional
flat indices. Upon compactification we split the indices according
to $M=(\mu,m)$, where $\mu, \nu, \dots$ and $m, n, \dots$ are curved
external and internal indices, respectively. Similarly, the flat
indices split into external and internal parts according to
$A=(\alpha,a)$.

Our reduction Ansatz for the vielbein is
\begin{equation}
\hvb{M}{A} = e^{\varphi} \tvb{M}{A} = e^{\varphi} \left(
\begin{array}{cc} \vb{\mu}{\alpha} & \mathcal{A}_{\mu}^m
\tvb{m}{a} \\
0 & \tvb{m}{a}\\
\end{array} \right),
\end{equation}
where the internal vielbein $\tvb{m}{a}$ is an element of the
isometry group $GL(n, \mbb{R})$ of the $n$-torus. Later on we shall
parametrise $\tvb{m}{a}$ in various ways. With this Ansatz,
the line element becomes
\beq
ds^2_D = e^{2\varphi} \Big\{ ds^2_{D-n} +
\big[(dx^m+\mathcal{A}_{(1)}^m)\tvb{m}{a}\big]^2 \Big\}.
\eeq

\subsubsection*{Weyl-Rescaling}

In order to obtain a Lagrangian in Einstein frame after dimensional
reduction, we perform a Weyl-rescaling of the $D$-dimensional
vielbein,
\beq
{\hat{e}_M}^{\ph{M}A}\hs \longrightarrow \hs
{\tilde{e}_M}^{\ph{M}A}=e^{-\varphi}{\hat{e}_M}^{\ph{M}A}.
\eeq

Note that all $D$-dimensional objects before rescaling are denoted
$\hat{X}$, the Weyl-rescaled objects are denoted $\tilde{X}$, while
the $d=(D-n)$-dimensional objects are written without any
diacritics. After the Weyl-rescaling the Gauss-Bonnet Lagrangian,
including the volume measure $\hat{e} = e^{D\varphi}\tilde{e}$, can
be conveniently organized in terms of equations of motion and total
derivatives. This is achieved using integration by parts, where
$\tilde{\nabla}_{(A}\tilde{\partial}_{B)}\varphi$ does not appear
explicitly. The resulting Lagrangian is (see Appendix
\ref{appendix}):
\beqa
{}Ê\mathcal{L}_{\mathrm{GB}} &=& \tilde{e} e^{(D-4)\varphi} \bigg\{
\tilde{R}_{\mathrm{GB}}^2 - (D-3)(D-4)\big[
2(D-2)(\tilde{\partial}\varphi)^2\tilde{\square}\varphi +
(D-2)(D-3)(\tilde{\partial}\varphi)^4 \nn \\
{}Ê& & + 4 \big(\tilde{R}_{AB} -
\frac{1}{2}\eta_{AB}\tilde{R}\big)
(\tilde{\partial}^A\varphi)(\tilde{\partial}^B\varphi)\big]\bigg\}
\nn \\
{}Ê& & + 2(D-3)\tilde{e}\tilde{\nabla}_A\bigg\{
e^{(D-4)\varphi}\big[(D-2)^2(\tilde{\partial}
\varphi)^2\tilde{\partial}^A\varphi + 2(D-2)(\tilde{\square}
\varphi)\tilde{\partial}^A\varphi
\nn \\
{}Ê& & - (D-2)\tilde{\partial}^A(\tilde{\partial} \varphi)^2 +
4\big(\tilde{R}^{AB} - \frac{1}{2}\eta^{AB}\tilde{R}\big)
\tilde{\partial}_B\varphi\big]\bigg\},
\eqnlab{eqn:GBweyl1}
\eeqa
where $\tilde{R}_{GB}^{2}$ represents the rescaled Gauss-Bonnet
combination. In $D=4$ the Lagrangian is only altered by a total
derivative, while in $D=3$ the Lagrangian it is merely rescaled by
a factor of $e^{-\varphi}$. The total derivative terms here will
remain total derivatives even after the compactification. Along
with the volume factor the Weyl-rescaling will determine the
overall exponential dilaton factor, which shall play an important
role in the analysis that follows.

\subsection{Tree-Level Scalar Coset Symmetries}

The internal vielbein $\hat{e}_m^{\phantom{m}a}$ can be used to
construct the internal metric $\hat{g}_{mn} =
\hat{e}_m^{\phantom{m}a}\hat{e}_n^{\phantom{m}b}\delta_{ab}$, which
is manifestly invariant under local $SO(n)$ rotations in the reduced
directions. Thus we are free to fix a gauge for the internal
vielbein using the $SO(n)$-invariance. After compactification the
volume measure becomes $\tilde{e}=e\tvbint$, where $e$ is the
determinant of the spacetime vielbein and $\tvbint$ is the
determinant of the internal vielbein. Defining the Weyl-rescaling
coefficient as $e^{-(D-2)\varphi} \equiv \tvbint$ ensures that the
reduced Lagrangian is in Einstein frame.

The $GL(n, \mbb{R})$ group element $\tvb{m}{a}$ can now be
parameterized in several ways, and we will discuss the two most
natural choices here. The first choice is included for completeness,
while it is the second choice which we shall subsequently employ in
the compactification of the Gauss-Bonnet term.

\subsubsection*{First Parametrisation - Making the Symmetry Manifest}

First, there is the possibility of separating out the determinant of
the internal vielbein according to $\tvb{m}{a} = (\tvbint)^{1/n}
\varepsilon_m^{\phantom{m}a} =
e^{-\frac{(D-2)}{n}\varphi}\varepsilon_m^{\phantom{m}a}$, where
$\varepsilon_m^{\phantom{m}a}$ is an element of $SL(n, \mbb{R})$ in
any preferred gauge. The line element takes the form
\beq
ds^2_D = e^{2\varphi} \left\{ ds^2_{D-n} +
e^{-2\frac{(D-2)}{n}\varphi} \left[(dx^m +
\mathcal{A}_{(1)}^m)\varepsilon_m^{\phantom{m}a}\right]^2 \right\}.
\eeq
This Ansatz is nice for investigating the symmetry properties of the
reduced Lagrangian because the $GL(n, \mbb{R})$-symmetry of the
internal torus is manifestly built into the formalism. More
precisely, the reduction of the tree-level Einstein-Hilbert
Lagrangian, $\hat{e}\hat{R}$, to $d=D-n$ dimensions becomes,
\beq
\mathcal{L}^{[d]}_{\mathrm{EH}} = e\left[ R -
\frac{1}{4}e^{-2\frac{(D-2)}{n}\xi\rho}
F_{c\alpha\beta}F^{c\alpha\beta} - \frac{1}{2}(\partial\rho)^2 -
\mathrm{tr}(P_\alpha P^\alpha) - 2\xi\square\rho \right],
\eqnlab{ReducedEH1}
\eeq
where $F^c_{\phantom{c}\alpha\beta} \equiv
\varepsilon_m^{\phantom{m}a} F^m_{\phantom{m}\alpha\beta}$ and
\beq
P_\alpha^{\phantom{\alpha}bc} \equiv \varepsilon^{m(b}
\partial_\alpha \varepsilon_m^{\phantom{m}c)} =
\tP_\alpha^{\phantom{\alpha} bc} +
\frac{(D-2)}{n}\xi\partial_\alpha\rho\delta^{bc}.
\eeq
Notice that $P_\alpha^{\phantom{\alpha} bc}$ is $\mf{sl}(n,
\mbb{R})$ valued and hence fulfills $\tr(P_\alpha) = 0$. To obtain
\Eqnref{ReducedEH1} we also performed a scaling $\varphi = \xi\rho$
with $\xi = \sqrt{\frac{n}{2(D-2)(D-n-2)}}$, so as to ensure that
the scalar field $\rho$ appears canonically normalized in the
Lagrangian.

The $SL(n, \mbb{R})$-symmetry is manifest in this Lagrangian because
the term $\mathrm{tr}(P_\alpha P^\alpha)$ is constructed using the
invariant Killing form on $\mf{sl}(n, \mbb{R})$. By dualising the
two-form field strength $F_{(2)}$, the symmetry is enhanced to
$SL(n+1, \mbb{R})$. With a slight abuse of terminology we call this
the (classical) ``U-duality'' group. Since we are only investigating
the pure gravity sector, this is of course only a subgroup of the
full continuous U-duality group.

It was already shown in \cite{Aspects}, that the tree-level symmetry
$SL(n+1, \mbb{R})$ is not realized in the compactified Gauss-Bonnet
Lagrangian. It was argued, however, that the quantum symmetry
$SL(n+1, \mbb{Z})$ could be reinstated by ``lifting'' the result of
the compactification through the use of automorphic forms. In this
paper we take the same point of view, but since we now have access
to the complete expression of the compactified Gauss-Bonnet
Lagrangian we can here extend the analysis of \cite{Aspects} in some
aspects. In order to do this we shall make use of a different
parametrisation than the one displayed above, which illuminates the
structure of the dilaton exponents in the Lagrangian. The dilaton
exponents reveals the weight structure of the global symmetry group
and so can give information regarding which representation of the
U-duality group we are dealing with. 

\subsubsection*{Second Parametrisation - Revealing the Root Structure}

The second natural choice of the internal vielbein is to
parameterize it in triangular form by using dimension by dimension
compactification \cite{PSolitons,Dualisation1,Dualisation2}. Instead
of extracting only the determinant of the vielbein, one dilaton
scalar is pulled out for each compactified dimension according to
$\tvb{m}{a} = e^{-\frac{1}{2}\fphi{a}}\ivb{m}{a}$, where $\vec{\phi}
= (\phi_1,\dots,\phi_n)$ and
\beq
\vec{f}_a = 2(\alpha_1,\dots,\alpha_{a-1},(D-n-2+a)\alpha_a,
\underbrace{0,\dots,0}_{n - a}),
\eqnlab{vectorf}
\eeq
with
\beq
\alpha_{a} = \frac{1}{\sqrt{2(D-n-2+a)(D-n-3+a)}}.
\eeq
The internal vielbein is now the Borel representative of the coset
$GL(n, \mbb{R})/SO(n)$, with the diagonal degrees of freedom
$e^{-\frac{1}{2}\fphi{a}}$ corresponding to the Cartan generators
and the upper triangular degrees of freedom
\beq
\ivb{m}{a} = [(1 - \mathcal{A}_{(0)})^{-1}]_m^{\phantom{m}a} = [1 +
\mathcal{A}_{(0)} + (\mathcal{A}_{(0)})^2 +
\dots]_m^{\phantom{m}a}
\eqnlab{uma}
\eeq
corresponding to the positive root generators. The form of
\Eqnref{uma} follows naturally from a step by step compactification,
where the scalar potentials $(\mathcal{A}_{(0)})^i_j$, arising from
the compactification of the graviphotons, are nonzero only when
$i>j$. The sum of the vectors $\vec{f}_a$ can be shown to be
\begin{equation}
\sum_{a=1}^n \vec{f}_a = \frac{D-2}{3}\vec{g},
\eqnlab{sumoff}
\end{equation}
$\vec{g} \equiv 6(\alpha_1,\alpha_2\dots,\alpha_n)$. In addition,
$\vec{g}$ and $\vec{f}_a$ obey
\beqa
{}Ê\vec{g}\cdot\vec{g} &=& \frac{18n}{(D-2)(D-n-2)},
\nn \\
{}Ê\vec{g}\cdot\vec{f}_a &= &\frac{6}{D-n-2},
\nn \\
{}Ê\vec{f}_a \cdot\vec{f}_b &= &2\delta_{ab} + \frac{2}{D-n-2},
\eqnlab{exponentrelations}
\eeqa
and
\beq
\sum_{a=1}^n (\vec{f}_a\cdot \vec{x})(\vec{f}_a\cdot \vec{y}) = 2
(\vec{x}\cdot\vec{y}) + \frac{D-2}{9}(\vec{g}\cdot
\vec{x})(\vec{g}\cdot\vec{y}).
\eeq
These scalar products can naturally be used to define the Cartan
matrix, once a set of simple root vectors are found. The line
element becomes
\beq
ds^2_D = e^{\frac{1}{3}\gphi} \Big\{ ds^2_{D-n} + \sum_{a=1}^n
e^{-\fphi{a}} \left[(dx^m+\mathcal{A}_{(1)}^m)\ivb{m}{a}\right]^2
\Big\},
\eqnlab{SecondAnsatz}
\eeq
yielding the corresponding Einstein-Hilbert Lagrangian in $d$
dimensions
\beq
\mathcal{L}_{\mathrm{EH}}^{[d]} = e\bigg[ R -
\frac{1}{2}(\partial\vec{\phi})^2 - \frac{1}{4}\sum_{a=1}^n
e^{-\fphi{a}}F_{a\beta\gamma}F^{a\beta\gamma} -
\frac{1}{2}\sum_{\substack{b,c=1 \\ b<c}}^n e^{(\vec{f}_b -
\vec{f}_c)\cdot\vec{\phi}}G_{\alpha bc} G^{\alpha bc} -
\frac{1}{3}\vec{g}\cdot\square\vec{\phi} \bigg],
\eeq
with $F^c_{\phantom{c}\alpha\beta} \equiv \ivb{m}{a}
F^m_{\phantom{m}\alpha\beta}$ and
\beq
G_\alpha^{\phantom{\alpha}bc} = u^{mb}\partial_\alpha\ivb{m}{c} =
e^{-\frac{1}{2}(\vec{f}_b -
\vec{f}_c)\cdot\vec{\phi}}\left[(\tP_\alpha^{\phantom{\alpha} bc} +
\frac{1}{2}\vec{f}_b\cdot\partial_\alpha \vec{\phi}\delta^{bc}) +
Q_\alpha^{\phantom{\alpha}bc}\right].
\eeq
Here, no Einstein's summation rule is assumed for the flat internal
indices. Notice also that $G_\alpha^{\phantom{\alpha}bc}$ is
non-zero only when $b<c$.

We shall refer to the various exponents of the form $e^{\vec{x}\cdot
\vp}$ ($\vec{x}$ being some vector in $\mbb{R}^{n}$) collectively as
``dilaton exponents''. If relevant, this also includes the
contribution from the overall volume factor.

All the results
obtained in this parametrisation can be converted to the first
parametrisation simply by using the following identifications
\beqa
{}Ê\frac{1}{3}(\gphi) &=& 2\xi\rho, \nn \\
{}Ê\fphi{a} &=& 2\frac{(D-2)}{n}\xi\rho, \hspace{0.5 cm} \forall a,
\nn \\
{}Ê\vscalar{\phi}{\phi} &=& \rho^2.
\eeqa
Notice also that our compactification procedure breaks down at $D-n
= 2$, in which case the scalar products in
\Eqnref{exponentrelations} become ill-defined.

Even though proving the symmetry contained in the Lagrangian is
somewhat more cumbersome compared to the first choice of
parametrisation, since all the group actions have to be carried out
adjointly in a formal manner, the second choice comes to its power
when dealing with the exceptional symmetry groups of the
supergravities for which no matrix representations exist. This
parametrisation is particularly suitable for reading off the root
vectors of the underlying symmetry algebra; they appear as
exponential factors in front of each term in the Lagrangian.
Identifying a complete set of root vectors in this way gives a
necessary but not sufficient constraint on the underlying symmetry.

\subsection{The Gauss-Bonnet Lagrangian Reduced to Three Dimensions}

When reducing to $D-n=3$ dimensions, we can dualise the two-form
field strength $\tF_{\phantom{a}\alpha\beta}^{a}\equiv
\tilde{e}_m^{\phantom{m}a}F^m_{\phantom{m}\alpha\beta}$ of the
graviphoton $\mc{A}_{(1)}$ into the one-form $\tH_{a \alpha}$. More
explicitly, we employ the standard dualisation
\beq
\delta_{ab}\tF^b_{\phantom{b}\alpha\beta} =
\epsilon_{\alpha\beta\gamma}\tilde{e}^m_{\phantom{m}a}
\partial^\gamma \chi_m \equiv \epsilon_{\alpha\beta\gamma}
\tH_a^{\phantom{a}\gamma}.
\eeq
When we go to Einstein frame, the appearance of the inverse vielbein
$\tilde{e}^m_{\phantom{m}a}$ in the definition of the one-form
$\tH_{a\alpha}$ implies there is a sign flip on its dilaton exponent
in the Lagrangian after dualisation. The dualisation presented here
follows from the tree-level Lagrangian, but in general receives
higher order $\al^{\prime}$-corrections. However, these lead to
terms of higher derivative order than quartic  and so can be
neglected in the present analysis \cite{LambertWest1,Aspects}.

The compactification is performed according to the standard
procedure by separating the indices; the detailed calculations can
be found in Appendix \ref{appendix}. The final results are written
in such way that the only explicit derivative terms appearing are
divergences, total derivatives and first derivatives on the dilatons
$\varphi$. The complete compactification of the Gauss-Bonnet Lagrangian on
$T^n$ to arbitrary dimensions $D-n$ is given in \Eqnref{GBtilde} of
Appendix \ref{appendix}.\footnote{Kaluza-Klein reduction of quadratic curvature corrections has also been analyzed from a different point of view in \cite{Jackiw}.} This expression is rather messy and
difficult to work with. However, by making use of all first order
equations of motion, dualising all graviphotons to scalars, and
restricting to $D-n=3$, the Lagrangian simplifies considerably. The
end result reads 
\beqa
{}Ê\mathcal{L}_{\mathrm{GB}}^{[3]}&=& \sqrt{|g|} e^{-2\varphi}
\Big\{ -\frac{1}{4}\tHHHH + \frac{1}{4}\tH^2 \tH^2 - 4\tH^2
(\partial\varphi)^2 + 2\tHPPHa  \nn \\
{}Ê& & - 2\tHPPHb + 4\tHPHa \partial_\beta \varphi - 6\tHPHb
\partial_\beta \varphi \nn \\
{}Ê& & + 2\tr(\tP_\alpha\tP_\beta \tP^\alpha\tP^\beta) +
2\tr(\tP_\alpha\tP_\beta) \tr(\tP^\alpha\tP^\beta) - (\tP^2)^2 +
8\tr(\tP_\alpha\tP_\beta \tP^\beta) \partial^\alpha\varphi \nn \\
{}Ê& &  - 4(D-2)\tr(\tP_\alpha
\tP_\beta)\partial^\alpha\varphi
\partial^\beta\varphi + 2(D+2)\tP^2 (\partial\varphi)^2 +
(D-2)(D-4)(\partial\varphi)^2(\partial\varphi)^2 \Big\},
\nn \\
\eqnlab{GB3}
\eeqa
where $\tH^2 \equiv \tH_{a\beta}\tH^{a\beta}$ and $\tP^2 \equiv
\tP_{\alpha bc}\tP^{\alpha bc}$. Note that contributions from the
boundary terms and terms proportional to the equations of motion
have been ignored. The one-form $\tilde{P}_{\al}$ is the
Maurer-Cartan form associated with the internal vielbein
$\tilde{e}_{m}^{\ph{a}a}$, and so takes values in the Lie algebra
$\mf{gl}(n, \mbb{R})=\mf{sl}(n, \mbb{R})\oplus \mbb{R}$. Here, the
abelian summand $\mbb{R}$ corresponds to the ``trace-part'' of
$\tilde{P}_{\al}$. Explicitly, we have
$\text{tr}(\tilde{P}_{\al})=-(D-2)\pa_{\al}\varphi$. We shall
discuss various properties of $\tilde{P}_{\al}$ in more detail
below.

Finally, we note that the three-dimensional Gauss-Bonnet term is
absent from the reduced Lagrangian because it vanishes identically
in three dimensions:
\beq
R_{\al\be\ga\de}R^{\al\be\ga\de} - 4R_{\al\be}R^{\al\be}+R^2=0,
\qquad (\al, \be, \ga, \de = 1, 2, 3).
\eeq
The remainder of this paper is devoted to a detailed analysis of the
symmetry properties of \Eqnref{GB3}.

\section{Algebraic Structure of the Compactified Gauss-Bonnet Term}

\label{section:AlgebraicStructure}

We have seen that the Ansatz presented in \Eqnref{SecondAnsatz} is
particularly suitable for identifying the roots of the relevant
symmetry algebra from the dilaton exponents associated with the
diagonal components of the internal vielbein. Through this analysis
one may deduce that for the lowest order effective action, the terms
in the action are organized according to the adjoint representation
of $\mf{sl}(n+1, \mbb{R})$, for which the weights are the roots. The
aim of this section is to extend the analysis to the Gauss-Bonnet
Lagrangian. By general arguments \cite{LambertWest1,LambertWest2},
it has been shown that the exponents no longer correspond to roots
of the symmetry algebra but rather they now lie on the weight
lattice. Here, however, we have access to the complete compactified
Lagrangian and we may therefore present an explicit counting of the
weights in the dilaton exponents and identify the relevant
$\mf{sl}(n+1, \mbb{R})$-representation.

An exhaustive analysis of the $\mf{sl}(4, \mbb{R})$-representation
structure of the Gauss-Bonnet term compactified from $6$ to $3$
dimensions on $T^3$ is performed. We do this in two alternative
ways.

First, we neglect the contribution from the overall dilaton factor
$e^{-2\varphi}$ in the representation structure. This is consistent
before dualisation because this factor is $SL(3,
\mbb{R})$-invariant. However, after dualisation this is no longer
true and one must understand what role this factor plays in the
algebraic structure. If one continues to neglect this factor then
all the weights fit into the ${\bf 84}$-representation of
$\mf{sl}(4, \mbb{R})$ with Dynkin labels $[2, 0, 2]$.

On the other hand, including the overall exponential dilaton factor
in the weight structure induces a shift on the weights so that the
highest weight is associated with the ${\bf 36}$-representation of
$\mf{sl}(4, \mbb{R})$ instead, with Dynkin labels $[2, 0, 1]$.
However, this representation is not ``big enough'' to incorporate
all the weights in the Lagrangian. It turns out that there are
additional weights outside of the ${\bf 36}$ that fit into a ${\bf
27}$-representation of $\mf{sl}(3, \mbb{R})$. Unfortunately there
seems to be no obvious argument for which
$\mf{sl}(4,\mbb{R})$-representation those ``extra'' weights should
belong to. 

This indicates that the first approach, where the dilaton pre-factor is neglected, is the correct way to interpret the result of the compactification because then all weights are ``unified'' in a single representation of the U-duality group. A detailed demonstration of this follows below.

\subsection{Kaluza-Klein Reduction and $\mf{sl}(n,
\mbb{R})$-Representations}

We shall begin by rewriting the reduction Ansatz in a way which has
a more firm Lie algebraic interpretation. Recall from
\Eqnref{SecondAnsatz} that the standard Kaluza-Klein Ansatz for the
metric is
\beq ds_D^{2}=e^{\f{1}{3}\vec{g}\cdot
\vec{\phi}}ds_{d}^{2}+e^{\f{1}{3}\gp}\sum_{i=1}^{n}e^{-\fip}
\left[\left(dx^{m} + \mc{A}^{m}_{(1)}\right){u_m}^{a}\right]^2.
\eqnlab{KaluzaKleinAnsatz}
\eeq
The exponents in this Ansatz are linear forms on the space of
dilatons. Let $\ve_i, \hs i=1, \dots, n$, constitute an
$n$-dimensional orthogonal basis of $\mbb{R}^{n}$, \beq
\ve_i\cdot\ve_j=\delta_{ij}. \eeq Since there is a non-degenerate
metric on the space of dilatons (the Cartan subalgebra
$\mf{h}\subset \mf{sl}(n+1, \mbb{R})$) we can use this to identify
this space with its dual space of linear forms. Thus, we may express
all exponents in the orthogonal basis $\ve_i$ and the vectors
$\vf_i, \vg$ and $\vp$ may then be written as

\beqa
{}Ê\vf_i& =&
\sqrt{2}\ve_i+\al\vg,
\nn \\
{}Ê\vg&=&\be \sum_{i=1}^{n}\ve_i,
\eqnlab{dilatonvectors}
\eeqa
where the constants $\al$ and $\be$ are defined as
\beqa
{}\al&=&\f{1}{3n}\bigg(D-2-\sqrt{(D-n-2)(D-2)}\bigg),
\nn \\
{}Ê  \be& =&\sqrt{\f{18}{(D-n-2)(D-2)}}.
\eeqa
Note here that the constant $\al$ is not the same as the $\al_a$ of
\Eqnref{vectorf}.

The combinations
\beq
\vf_i-\vf_j = \sqrt{2}\ve_i-\sqrt{2}\ve_j
\eeq
span an $(n-1)$-dimensional lattice which can be identified with the
root lattice of $A_{n-1}=\mf{sl}(n, \mbb{R})$. For compactification
of the pure Einstein-Hilbert action to three dimensions, the dilaton
exponents precisely organize into the complete set of positive roots
of $\mf{sl}(n, \mbb{R})$, revealing that it is the adjoint
representation which is the relevant one for the U-duality
symmetries of the lowest order (two-derivative) action. After
dualisation of the Kaluza-Klein one forms $\mc{A}_{(1)}$ the
symmetry is lifted to the full adjoint representation of
$\mf{sl}(n+1, \mbb{R})$.

When we compactify higher derivative corrections to the
Einstein-Hilbert action it is natural to expect that other
representations of $\mf{sl}(n, \mbb{R})$ and $\mf{sl}(n+1, \mbb{R})$
become relevant. In order to pursue this question for the
Gauss-Bonnet Lagrangian, we shall need some features of the
representation theory of $\mf{sl}(n+1, \mbb{R})$.

\subsubsection*{Representation Theory of $A_n=\mf{sl}(n+1, \mbb{R})$}

For the infinite class of simple Lie algebras $A_n$, it is possible
to choose an embedding of the weight space $\mf{h}^{\star}$ in
$\mbb{R}^{n+1}$ such that $\mf{h}^{\star}$ is isomorphic to the
subspace of $\mbb{R}^{n+1}$ which is orthogonal to the vector
$\sum_{i=1}^{n+1}\ve_i$ (see, e.g., \cite{FS}). We can use this fact
to construct an embedding of the $(n-1)$-dimensional weight space of
$A_{n-1}=\mf{sl}(n, \mbb{R})$ into the $n$-dimensional weight space
of $A_n=\mf{sl}(n+1, \mbb{R})$, in terms of the $n$ basis vectors
$\ve_i$ of $\mbb{R}^{n}$.

To this end we define the new vectors
\beqa
{}Ê \vo_i &=&
\vf_i-(\al+\f{\sqrt{2}}{n\be})\vg
\nn \\
{}Ê& =& \sqrt{2}\ve_i-\f{\sqrt{2}}{n}\sum_{j=1}^{n}\ve_j,
\eeqa
which have the property that
\beq
\vo_i\cdot \vg
=\sqrt{2}\be-\sqrt{2}\be=0.
\eqnlab{zeroscalarproduct}
\eeq
This implies that the vectors $\vo_i$ form a (non-orthogonal) basis
of the $(n-1)$-dimensional subspace $U\subset \mbb{R}^{n}$,
orthogonal to $\vg$. The space $U$ is then isomorphic to the weight
space $\mf{h}^{\star}$ of $A_{n-1}=\mf{sl}(n, \mbb{R})$. Since there
are $n$ vectors $\vo_i$, this basis is overcomplete. However, it is
easy to see that not all $\vo_i$ are independent, but are subject to
the relation
\beq
\sum_{i=1}^{n}\vo_i=0.
\eqnlab{tracelessness}
\eeq
A basis of simple roots of $\mf{h}^{\star}$ can now be written in
three alternative ways
\beq
\vec{\al}_i=\vf_i-\vf_{i+1}=\vo_i-\vo_{i+1}=\sqrt{2}(\ve_i-\ve_{i+1}),
\qquad (i=1, \dots, n-1).
\eeq

What is the algebraic interpretation of the vectors $\vo_i$? It
turns out that they may be identified with the weights of the
$n$-dimensional fundamental representation of $\mf{sl}(n, \mbb{R})$.
The condition $\sum_{i=1}^{n}\vo_i=0$ then reflects the fact that
the generators of the fundamental representation are traceless.

In addition, we can use the weights of the fundamental
representation to construct the fundamental weights
$\vec{\Lambda_i}$, defined by
\beq
\vec{\al}_i\cdot \vec{\Lambda}_j=2\delta_{ij}.
\eqnlab{fundamentalweightsdefinition}
\eeq
One finds
\beq
\vL_i=\sum_{j=1}^{i}\vo_j, \qquad (i=1, \dots, n-1),
\eqnlab{weightrelation}
\eeq
which can be seen to satisfy \Eqnref{fundamentalweightsdefinition}.

The relation, \Eqnref{weightrelation}, between the fundamental
weights $\vL_i$ and the weights of the fundamental representation
$\vo_i$ can be inverted to
\beq
\vo_i = \vL_i-\vL_{i-1}, \qquad (i=1, \dots, n-1).
\eeq
In addition, the $n$:th weight is
\beq \vo_n=-\vL_{n-1},
\eqnlab{nthWeight}
\eeq
corresponding to the lowest weight of the fundamental
representation.

We may now rewrite the Kaluza-Klein Ansatz in a way such that the
weights $\vo_i$ appear explicitly in the metric\footnote{A similar
construction was given in \cite{Arjan}.}
\beq
ds_D^{2} = e^{\f{1}{3}\vec{g}\cdot
\vec{\phi}}ds_{d}^{2}+e^{\ga\gp}\sum_{i=1}^{n}e^{-\vo_i\cdot
\vp}\Big[\big(dx^{m}+\mc{A}^{m}_{(1)}\big){u_m}^{a}\Big]^2,
\eqnlab{KaluzaKleinAnsatz2}
\eeq
with
\beq
\ga = \f{1}{3} - \al-\f{\sqrt{2}}{n\be}.
\eeq

\subsection{The Algebraic Structure of Gauss-Bonnet in Three
Dimensions}

We are interested in the dilaton exponents in the scalar part of the
three-dimensional Lagrangian. For the Einstein-Hilbert action we
know that these are of the forms
\beq
Ê\vf_a-\vf_b \quad (b>a), \qquad \text{and}\qquad \vf_a.
\eeq
The first set of exponents $\vf_a-\vf_b$ correspond to the positive
roots of $\mf{sl}(n, \mbb{R})$ and the second set $\vf_a$, which
contributes to the scalar sector after dualisation, extends the
algebraic structure to include all positive roots of $\mf{sl}(n+1,
\mbb{R})$. The highest weight $\vl^{\text{hw}}_{\text{ad}, n}$ of
the adjoint representation of $A_n=\mf{sl}(n+1, \mbb{R})$ can be
expressed in terms of the fundamental weights as
\beq
\vl^{\text{hw}}_{\text{ad}, n}=\vL_{1}+\vL_{n},
\eeq
corresponding to the Dynkin labels $$[1, 0, \dots, 0, 1].$$ We see
that before dualisation the highest weight of the adjoint
representation of $\mf{sl}(n, \mbb{R})$ arises in the dilaton
exponents in the form
$\vf_1-\vf_n=\vo_1-\vo_n=\vL_1+\vL_{n-1}=\vl^{\text{hw}}_{\text{ad},
n-1}$.

We proceed now to analyze the various dilaton exponents arising from
the Gauss-Bonnet term after compactification to three dimensions.
These can be extracted from each term in the Lagrangian \Eqnref{GB3}
by factoring out the diagonal components of the internal vielbein
according to $\tvb{m}{a} = e^{-\frac{1}{2}\fphi{a}}\ivb{m}{a}$. For
example, before dualisation we have the manifestly $SL(n,
\mbb{R})$-invariant term $\text{tr}(\tilde{P}_{\al}\tilde{P}_{\be}
\tilde{P}^{\al}\tilde{P}^{\be})$. Expanding this gives (among
others) the following types of terms
\beqa
{}Ê\text{tr}\big(\tilde{P}_{\al}\tilde{P}_{\be}
\tilde{P}^{\al}\tilde{P}^{\be}\big) & \sim & \ph{+}
\sum_{\tiny\begin{array}{l}b < a,c \\ d < a,c
\end{array}} e^{-(\vf_a-\vf_b+\vf_c-\vf_d)\cdot \vp} G_{\al b
a} G_{\be}^{\phantom{\be}bc} G^{\al}_{\phantom{\al}dc} G^{\be da} +
\cdots \nn \\
{}Ê & & + \sum_{\tiny\begin{array}{l}a<c<d<b\end{array}}
e^{(\vf_a-\vf_b)\cdot \vp} G_{\al ab} G_{\be}^{\phantom{\be}ac}
G^{\al}_{\phantom{\al}cd} G^{\be db} +\cdots.
\eqnlab{IndexExtraction}
\eeqa
After dualisation, we need to take into account also terms
containing $\tilde{H}^{\al}_{\ph{a} a}$. We have then, for example,
the term
\beq
\tilde{H}^4 \hs \sim \hs \sum_{a, b}e^{(\vf_a+\vf_b)\cdot \vp} H^4.
\eeq
Many different terms in the Lagrangian might in this way give rise
to the same dilaton exponents. As can be seen from
\Eqnref{IndexExtraction}, the internal index contractions yield
constraints on the various exponents. We list below all the
``independent'' exponents, i.e., those which are the least
constrained. All other exponents follow as special cases of these.
Before dualisation we find the following exponents:
\beqa
{}Ê\vf_a-\vf_b &\quad & (b>a),
\nn \\
{}Ê\vf_c+\vf_d-\vf_a-\vf_b &\quad & (c<a,\hs c<b, \hs d<a, \hs d<b),
\nn \\
{}Ê\vf_a+\vf_b-\vf_c-\vf_d & \quad & (b<c, \hs a<d),
\eqnlab{GBexponents1}
\eeqa
and after dualisation we also get contributions from
\beqa
{} \vf_a, & &
\nn \\
{}Ê\vf_a+\vf_b, &Ê&
\nn \\
{}Ê\vf_a+\vf_b-\vf_c, &\quad & (a<c,\hs b<c).
\eqnlab{GBexponents2}
\eeqa
Let us first investigate the general weight structure of the dilaton
exponents before dualisation. The highest weight arises from the
terms of the form $\vf_c+\vf_d-\vf_a-\vf_b$ when $c=d=1$ and
$a=b=n$, i.e., for the dilaton vector $2\vf_1-2\vf_n$. This can be
written in terms of the fundamental weights as follows
\beq
2\vf_1-2\vf_n=2\vo_1-2\vo_n=2\vL_1+2\vL_{n-1},
\eeq
which is the highest weight of the $[2, 0, \dots, 0,
2]$-representation of $\mf{sl}(n, \mbb{R})$.

\subsection{Special Case: Compactification from $D=6$ on $T^{3}$}

\label{section:SpecialCaseNoVolumeFactor}

In order to determine if this is indeed the correct representation
for the Gauss-Bonnet term, we shall now restrict to the case of
$n=3$, i.e., compactification from $D=6$ on $T^{3}$. We do this so
that a complete counting of the weights in the Lagrangian is a
tractable task. Before dualisation we then expect to find the
representation ${\bf 27}$ of $\mf{sl}(3, \mbb{R})$, with Dynkin
labels $[2,2]$. We will see that, after dualisation, this
representation lifts to the representation ${\bf 84}$ of $\mf{sl}(4,
\mbb{R})$, with Dynkin labels $[2, 0, 2]$.

It is important to realize that of course the Lagrangian will not
display the complete set of weights in these representations, but
only the \emph{positive} weights, i.e., the ones that can be
obtained by summing positive roots only. Let us begin by analyzing
the weight structure before dualisation. From \Eqnref{GBexponents1}
we find the weights
\beqa
{}Ê& & \vf_1-\vf_2, \qquad \vf_2-\vf_3, \qquad \vf_1-\vf_3,
\nn \\
{}Ê& & 2(\vf_1-\vf_2), \qquad 2(\vf_2-\vf_3), \qquad 2(\vf_1-\vf_3),
\nn \\
{}Ê& & 2\vf_1-\vf_2-\vf_3, \qquad \vf_1+\vf_2-2\vf_3.
\eeqa
The first three may be identified with the positive roots of
$\mf{sl}(3, \mbb{R})$, $\val_1=\vf_1-\vf_2,\hs \val_2=\vf_2-\vf_3$
and $\val_{\theta}=\vf_1-\vf_3$. The second line then corresponds to
$2\val_2,\hs 2\val_2$ and $2\val_{\theta}$. The remaining weights
are
\beqa
{}Ê\vf_1+\vf_2-2\vf_3&=&\val_1+2\val_2,
\nn \\
{}Ê 2\vf_1-\vf_2-\vf_3&=&2\val_1+\val_2.
\eeqa
These weights are precisely the eight positive weights of the {\bf
27} representation of $\mf{sl}(3, \mbb{R})$.

We now wish to see whether this representation lifts to any
representation of $\mf{sl}(4, \mbb{R})$, upon inclusion of the
weights in \Eqnref{GBexponents2}. As mentioned above, the natural
candidate is an 84-dimensional representation of $\mf{sl}(4,
\mbb{R})$ with Dynkin labels $[2, 0, 2]$. It is illuminating to
first decompose it in terms of representations of $\mf{sl}(3,
\mbb{R})$,
\beq
{\bf 84}={\bf 27}\oplus {\bf 15}\oplus {\bf \bar{15}}\oplus {\bf
6}\oplus {\bf \bar{6}}\oplus {\bf 8}\oplus {\bf 3}\oplus {\bf
\bar{3}}\oplus {\bf 1},
\eqnlab{84inSL3}
\eeq
or, in terms of Dynkin labels,
\beq
[2, 0, 2]=[2,2]+[2,1]+[1,2]+[2,0]+[0,2]+[1,1]+[1,0]+[0,1]+[0,0].
\eeq
We may view this decomposition as a \emph{level decomposition} of
the representation ${\bf 84}$, with the level $\ell$ being
represented by the number of times the third simple root $\val_3$
appears in each representation. From this point of view, and as we
shall see in more detail shortly, the representations ${\bf 27},
{\bf 8}$ and ${\bf 1}$ reside at $\ell=0$, the representations ${\bf
15}$ and ${\bf 3}$ at $\ell=1$, and the representation ${\bf 6}$ at
$\ell=2$. The ``barred'' representations then reside at the
associated negative levels. Knowing that we will only find the
strictly positive weights in these representations, let us therefore
start by listing these.

\begin{figure}
\begin{center}
\includegraphics[width=150mm]{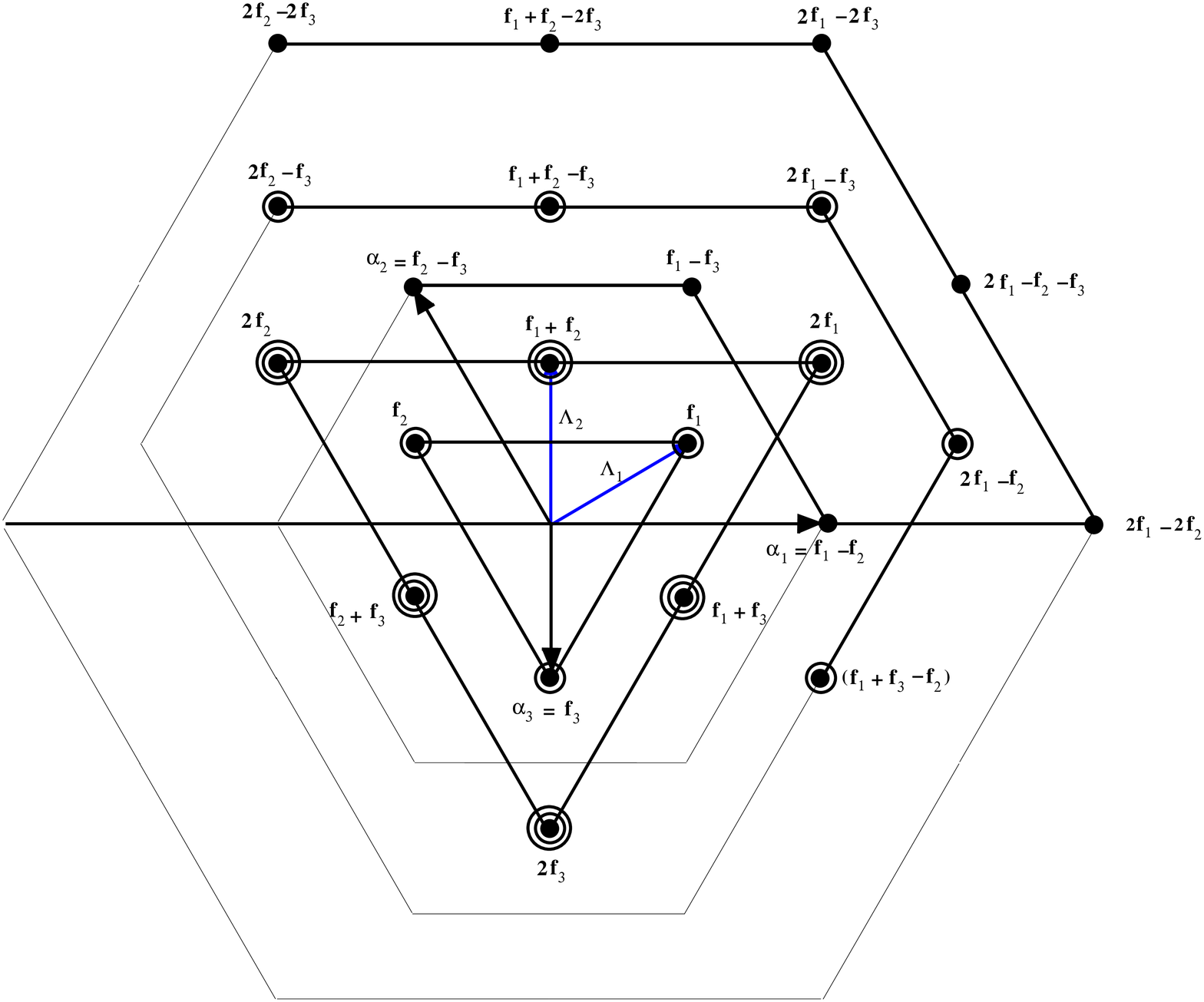}
\caption{Graphical presentation of the representation structure of
the compactified Gauss-Bonnet term. The black nodes arise from
distinct dilaton exponents in the three-dimensional Lagrangian. The
figure displays the level decomposition of the ${\bf
84}$-representation of $\mf{sl}(4, \mbb{R})$ into representations of
$\mf{sl}(3, \mbb{R})$. Only positive levels are displayed. The black
nodes correspond to positive weights of ${\bf 84}$ of $\mf{sl}(4,
\mbb{R})$. Nodes with no rings represent the positive weights of the
level zero representation ${\bf 27}$, nodes with one ring represent
the positive weights of the level one representations ${\bf 15}$ and
${\bf 3}$, while nodes with two rings represent the positive weights
of the level two representation ${\bf 6}$. The shaded lines complete
the representations with non-positive weights which are not
displayed explicitly. The missing weight is put into a parenthesis.}
\label{figure:84ofSL4}
\end{center}
\end{figure}

Firstly, we may neglect all representations at negative levels since
these do not contain any positive weights. However, not all weights
for $\ell\geq 0$ are positive. If we had decomposed the adjoint
representation of $\mf{sl}(4, \mbb{R})$ this problem would not have
been present since all roots are either positive or negative, and
hence all weights at positive level are positive and vice versa. In
our case this is not true because for representations larger than
the adjoint many weights are neither positive nor negative. It is
furthermore important to realize that after dualisation it is the
positive weights of $\mf{sl}(4, \mbb{R})$ that we will obtain and
not of $\mf{sl}(3, \mbb{R})$. As can be seen in Figure
\ref{figure:84ofSL4} the decomposition indeed includes weights which
are negative weights of $\mf{sl}(3, \mbb{R})$ but nevertheless
positive weights of $\mf{sl}(4, \mbb{R})$. An explicit counting
reveals the following number of positive weights at each level (not
counting weight multiplicities):
\beqa
{}Ê\ell=0 &:&  8,
\nn \\
{}Ê\ell=1 &:& 8,
\nn \\
{}Ê\ell=2 &:& 6.
\eeqa
The eight weights at level zero are of course the positive weights
of the ${\bf 27}$ representation of $\mf{sl}(3, \mbb{R})$ that we
had before dualisation. In order to verify that we find all positive
weights of ${\bf 84}$ we must now check explicitly that after
dualisation we get $8+6$ additional positive weights. The total
number of distinct weights of $\mf{sl}(4, \mbb{R})$ that should
appear in the Lagrangian after compactification and dualisation is
thus 22.

The lifting from $\mf{sl}(3, \mbb{R})$ to $\mf{sl}(4, \mbb{R})$ is
done by adding the third simple root $\val_3\equiv \vf_3$, from
\Eqnref{GBexponents2}. The complete set of new weights arising from
\Eqnref{GBexponents2} is then
\beqa
{}Ê\ell=1 &:&  \vf_1=\val_1+\val_2+\val_3, \qquad \qquad \quad
\hs\vf_2=\val_2+\val_3,
\nn \\
{}Ê& & 2\vf_1-\vf_2=2\val_1+\val_2+\val_3, \qquad
2\vf_2-\vf_3=2\val_2+\val_3,
\nn \\
{}Ê& & 2\vf_1-\vf_3=2\val_1+2\val_2+\val_3, \qquad
\vf_1+\vf_2-\vf_3=\val_1+2\val_2+\val_3,
\nn \\
{}Ê& & (\vf_1+\vf_3-\vf_2=\val_1+\val_3), \qquad \quad \vf_3=\val_3,
\nn \\
\nn \\
{}Ê\ell=2 &:& 2\vf_1=2\val_1+2\val_2+2\val_3, \qquad \quad \hs
2\vf_2=2\val_2+2\val_3,
\nn \\
{}Ê& & 2\vf_3=2\val_3, \qquad \qquad \qquad \quad
\qquad\vf_1+\vf_2=\val_1+2\val_2+2\val_3,
\nn \\
{}Ê& & \vf_1+\vf_3=\val_1+\val_2+2\val_3, \qquad \quad
\vf_2+\vf_3=\val_2+2\val_3. \eeqa In Table \ref{table:Reps} we
indicate which representations these weights belong to and in
Figure \ref{figure:84ofSL4} we give a graphical presentation of
the level decomposition. The weight $\val_1+\val_3$ is put inside
a parenthesis since terms giving this particular dilaton exponent
in the Gauss-Bonnet combination are all absorbed into the
equations of motion, and thus do not contribute according to our
compactification procedure. However, generically it will
contribute for a general second order curvature correction. We
suspect the origin of this ``missing'' weight is connected to the
mismatch in the multiplicity counting, which we will discuss
briefly below. These results show that the Gauss-Bonnet term in
$D=6$ compactified on $T^3$ to three dimensions gives rise to
strictly positive weights that can all be fit into the ${\bf
84}$-representation of $\mf{sl}(4, \mbb{R})$.

\begin{table}
\begin{center}
\begin{tabular}{|m{15mm}|m{15mm}|m{80mm}|}
\hline
& &  \\
Reps & $\ell$ & Positive Weights of $\mf{sl}(4, \mbb{R})$ \\
& &  \\
\hline
& & \\
${\bf 3}$ & 1Ê& $\val_3, \hs \val_2+\val_3, \hs
\val_1+\val_2+\val_3$ \\
& & \\
\hline
& & \\
${\bf 15}$ & 1 & $2\val_2+\val_3, \hs \val_1+2\val_2+\val_3, \hs
2\val_1+2\val_2+\val_3$, \\
& & \\
& & $2\val_1+\val_2+\val_3, \hs (\val_1+\val_3)$ \\
& & \\
\hline
& & \\
$ {\bf 6}$ & 2 & $2\val_2+2\val_3, \hs \val_1+2\val_2+2\val_3, \hs
2\val_1+2\val_2+2\val_3$, \\
 & & \\
 & & $\val_1+\val_2+2\val_3, \hs 2\val_3, \hs \val_2+2\val_3$ \\
 & & \\
\hline
\end{tabular}
\caption{Positive weights at levels one and two.}
\label{table:Reps}
\end{center}
\end{table}

\subsubsection*{Weight Multiplicities}

We have shown that the six-dimensional Gauss-Bonnet term
compactified to three dimensions gives rise to positive weights of
the ${\bf 84}$-representation of $\mf{sl}(4, \mbb{R})$. However, we
have not yet addressed the issue of weight multiplicities. It is not
clear how to approach this problem. Naively, one might argue that if
$k$ distinct terms in the Lagrangian are multiplied by the same
dilaton exponential, corresponding to some weight $\vec{\lambda}$,
then this weight has multiplicity $k$. Unfortunately, this type of
counting does not seem to work, one of the reasons being that the
notion of distinctness is not clearly defined.

Consider, for instance, the representations at $\ell=1$. Both
representations ${\bf 15}$ and ${\bf 3}$ contain the weights $\vf_1,
\hs \vf_2$ and $\vf_3$. In ${\bf 15}$ these have all multiplicity 2,
while in ${\bf 3}$ they have multiplicity 1. Thus, in total these
weights have multiplicity 3 as weights of $\mf{sl}(3, \mbb{R})$.
Now, a detailed investigation reveals that the dilaton exponent
$\vf_a$ appears in the Gauss-Bonnet term accompanied with various
different constraints on the index $a$, the no constraint case given
in \Eqnref{GBexponents2} is merely the ``most unconstrained'' one.
It can be easily shown that weights with lower value on index $a$
have higher multiplicity. We therefore deduce that for all these
weights there appears to be a mismatch in the multiplicity.

We suggest that the correct way to interpret this discrepancy in the
weight multiplicities is as an indication of the need to introduce
transforming automorphic forms in order to restore the $SL(4,
\mbb{Z})$-invariance. This will be discussed more closely in Section
\ref{section:resolution}.

\subsubsection*{Including the Dilaton Prefactor}

\label{section:SpecialCaseWithVolumeFactor}

We will now revisit the analysis from Section
\ref{section:SpecialCaseNoVolumeFactor}, but here we include the
contribution from the overall exponential factor $e^{-2\varphi}$ in
the Lagrangian \Eqnref{GB3}. This factor arises as follows. The
determinant of the $D$-dimensional vielbein is given by
$\hat{e}=e^{D\varphi}\tilde{e}$, because of the Weyl-rescaling.
Moreover, upon compactification the determinant of the rescaled
vielbein splits according to $\tilde{e}=e \tilde{e}_{\text{int}}$,
where $e$ represents the external vielbein and
$\tilde{e}_{\text{int}}$ the internal vielbein. The Weyl-rescaling
is then chosen to be defined as
$\tilde{e}_{\text{int}}=e^{-(D-2)\varphi}$. This represents the
volume of the $n$-torus, upon which we perform the reduction. Thus,
the overall scaling contribution from the measure is
$e^{D\varphi}e^{-(D-2)\varphi}=e^{2\varphi}$. In addition, we have a
factor of $e^{-4\varphi}$ from Weyl-rescaling the Gauss-Bonnet term
(see \Eqnref{RiemannSquared} and \Eqnref{RicciSquared}). This gives
a total overall dilaton prefactor of $e^{-2\varphi}$, which, after
inserting $\varphi=\f{1}{6}\vg \cdot \vp$, becomes $e^{-\f{1}{3}\vg
\cdot \vp}$.

The importance of the volume factor for compactified higher
derivative terms was emphasized in \cite{LambertWest1}, using the
argument that after dualisation this factor is no longer invariant
under the extended symmetry group $SL(n+1, \mbb{R})$ and so must be
included in the weight structure. We shall see that the inclusion of
this factor drastically modifies the previously presented structure.

\vspace{.6cm}

\noi {\small \it \textbf{The Fundamental Weights of $\mf{sl}(4,
\mbb{R})$}}

\vspace{.2cm}

\noi In order to perform this analysis, it is useful to first
rewrite the simple roots and fundamental weights in a way which
makes a comparison with \cite{LambertWest1} possible. We define
arbitrary $3$-vectors in $\mbb{R}^3$ as follows
\beq
\hat{\vec{v}}=v_1\vL_1+v_2\vL_2+v_g\vg = \big(\vec{v},
v_g\big)=\big(v_1, v_2, v_g\big),
\eeq
where $\vL_1$ and $\vL_2$ are the fundamental weights of $\mf{sl}(3,
\mbb{R})$ and $\vg$ is the basis vector taking us from the weight
space $\mbb{R}^2$ of $\mf{sl}(3, \mbb{R})$ to the weight space
$\mbb{R}^3$ of $\mf{sl}(4, \mbb{R})$. Note that
\beq
\vL_1\cdot \vg=\vL_2\cdot \vg=0,
\eeq
by virtue of \Eqnref{zeroscalarproduct} and \Eqnref{weightrelation},
which implies
\beq
\hat{\vec{v}}\cdot \hat{\vec{u}}=\vec{v}\cdot \vec{u} + v_g u_g \vg
\cdot \vg.
\eeq

The scalar products may all be deduced using the orthonormal basis
$\ve_i$ of $\mbb{R}^3$. Restricting to $D=6$ and $n=3$ gives
\beq
\vf_a=\sqrt{2}\ve_a+\f{2}{9}\vg,
\eeq
and thus
\beq
\vo_a=\vf_a-\f{4}{9}\vg.
\eeq
The relevant scalar products become
\beqa
{}Ê\vg\cdot \vg &=& \f{27}{2},
\nn \\
{}Ê\vg\cdot \vf_a&=&6,
\nn \\
{}Ê\vf_a\cdot \vf_b&=& 2\delta_{ab}+2,
\nn \\
{}Ê\vo_a\cdot \vo_b&=& 2\delta_{ab}-\f{2}{3}.
\eeqa
The simple roots of $\mf{sl}(3, \mbb{R})$ may now be written as
\beqa
{}Ê\hat{\val}_1&=& \big(\val_1, 0\big)=\big(2, -1, 0\big),
\nn \\
{}Ê\hat{\val}_2 &=& \big(\val_2, 0\big)=\big(-1, 2, 0\big),
\eeqa
and the third simple root becomes
\beq
\hat{\val}_3= \vf_3=\vo_3+\f{4}{9}\vg=-\vL_2+\f{4}{9}\vg=\big(0, -1,
\f{4}{9}\big).
\eeq
In addition, the associated fundamental weights $\hat{\vL}_i, \hs
i=1, 2, 3, $ of $\mf{sl}(4, \mbb{R})$, defined by
\beq
\hat{\val}_i\cdot \hat{\vL}_j=2\delta_{ij},
\eeq
become
\beq \hat{\vL}_1=\big(1, 0, \f{1}{9}\big),\quad \hat{\vL}_2 = \big(0,
1, \f{2}{9}\big),\quadÊ\hat{\vL}_3 = \big(0,0, \f{1}{3}\big).
\eeq
Let us check that these indeed correspond to the fundamental weights
of $\mf{sl}(4, \mbb{R})$, by computing the highest weight
$2\hat{\vL}_1+2\hat{\vL}_3$ explicitly,
\beqa
{}Ê2\hat{\vL}_1+2\hat{\vL}_3&=&2\vL_1+\f{2}{9}\vg+\f{2}{3}\vg
\nn \\
{}Ê&= &2(\vo_1+\f{4}{9}\vg)
\nn \\
{}Ê& =&2\vf_1
\nn \\
{}Ê&=&2\hat{\val}_1+2\hat{\val}_2+2\hat{\val}_3.
\eeqa
This result is consistent with being the highest weight of the ${\bf
84}$ representation of $\mf{sl}(4, \mbb{R})$ as can be seen in
Figure \ref{figure:84ofSL4}.

\vspace{.3cm}

\noi {\small \it \textbf{Dualisation and the Overall Dilaton Factor}}

\vspace{.2cm}

\noi Let us now include the dilaton prefactor in the analysis. In
terms of $\mf{sl}(4, \mbb{R})$-vectors the volume factor can be
identified with a negative shift in $\hat{\vL}_3$, i.e.,
\beq
e^{-\f{1}{3}\vg\cdot \vp}=e^{-\hat{\vL}_3\cdot \vp}.
\eeq
As already mentioned above, this factor is irrelevant before
dualisation because $\vg\cdot \vp$ is invariant under $SL(3,
\mbb{R})$. Thus, before dualisation the manifest $SL(3,
\mbb{R})$-symmetry of the compactified Gauss-Bonnet term is
associated with the ${\bf 27}$-representation of $\mf{sl}(3,
\mbb{R})$.

After dualisation, all the dilaton exponents in
\Eqnref{GBexponents1} and \Eqnref{GBexponents2} become shifted by a
factor of $-\hat{\vL}_3$. In particular, the new highest weight is
\beq
\big(2\hat{\vL}_1+2\hat{\vL}_3\big)-\hat{\vL}_3=2\hat{\vL}_1 +
\hat{\vL}_3,
\eeq
corresponding to the ${\bf 36}$ representation of $\mf{sl}(4,
\mbb{R})$, with Dynkin labels $[2, 0, 1]$. This is consistent with
the general result of \cite{LambertWest1} that a generic curvature
correction to pure Einstein gravity of order $l/2$ should be
associated with an $\mf{sl}(n+1, \mbb{R})$-representation with
highest weight $\f{l}{2}\hat{\vL}_1+\hat{\vL}_n$.

However, this is not the full story. A more careful examination in
fact reveals that the ${\bf 36}$ representation cannot incorporate
all the dilaton exponents appearing in the Lagrangian, in contrast
to the ${\bf 84}$-representation of Figure \ref{figure:84ofSL4}. To
see this, let us decompose ${\bf 36}$ in terms of representations of
$\mf{sl}(3, \mbb{R})$. The result is:
\beqa
{}Ê{\bf 36} &=& {\bf 15}\oplus {\bf 8}\oplus {\bf 6}\oplus {\bf
3}\oplus {\bf \bar{3}}\oplus {\bf 1},
\nn \\
{}Ê[2, 0, 1]&=& [2,1] + [1,1] + [2,0] + [1,0] + [0,1] + [0,0].
\eqnlab{36inSL3} \eeqa Comparing this with \Eqnref{84inSL3}, we
see that the representations ${\bf 27}, {\bf \bar{15}}$ and ${\bf
\bar{6}}$ are no longer present. For the latter two this is not a
problem since they were never present in the previous analysis.
What happens is that the ${\bf 6}$ of ${\bf 84}$ gets shifted
``downwards'' and becomes the ${\bf 6}$ of ${\bf 36}$. Similarly,
the ${\bf 15}$ and ${\bf 3}$ of ${\bf 84}$ become the ${\bf 15}$
and ${\bf 3}$ of ${\bf 36}$. This takes into account all the
shifted dilaton exponents arising from the dualisation process.
However, since there is not enough ``room'' for the ${\bf 27}$ of
$\mf{sl}(3, \mbb{R})$ in \Eqnref{36inSL3}, some of the dilaton
exponents (the ones corresponding to $2\vf_2-2\vf_3, \hs
\vf_1+\vf_2-2\vf_3, \hs 2\vf_1-2\vf_3, 2\vf_1-\vf_2-\vf_3$ and
$2\vf_1-2\vf_2$) arising from the pure $\tilde{P}$-terms remain
outside of ${\bf 36}$. In fact, due to the shift of $-\hat{\vL}_3$
these have now become \emph{negative} weights of $\mf{sl}(4,
\mbb{R})$, because they are below the hyperplane defined by
$\vg\cdot \vec{x}=0$. Although we know that these weights still
correspond to positive weights of the ${\bf 27}$ representation of
$\mf{sl}(3, \mbb{R})$, we are not able to determine which
representation of $\mf{sl}(4, \mbb{R})$ they belong to.

By a straightforward generalisation of this analysis to
compactifications of quadratic curvature corrections from arbitrary
dimensions $D$, we may conclude that the highest weight
$2\hat{\vL}_1+\hat{\vL}_n$, can never incorporate the dilaton
exponents associated with the $[2, 0,$ $ \dots, 0,
2]$-representation of $\mf{sl}(n, \mbb{R})$ before dualisation.

\section{Discussion and Conclusions}
\label{section:resolution}

It is clear from the analysis in the previous section that the
overall dilaton factor $e^{-\hat{\vL}_3\cdot \vp}$ (or, more
generally, $e^{-\hat{\vL}_n\cdot \vp}$) complicates the
interpretation of the dilaton exponents in terms of $\mf{sl}(n+1,
\mbb{R})$-representations. A similar problem has arisen in attempts
at incorporating the representation structure of the hyperbolic
Kac-Moody algebra $E_{10(10)}$ into curvature corrections to string
and M-theory \cite{DamourNicolai,DHHKN}. There it is the ``lapse
function'' $N$ which plays the role of the volume factor. Similarly
to our findings, the work of \cite{DamourNicolai,DHHKN} reveals that
curvature corrections to, e.g., eleven-dimensional supergravity, fit
into \emph{negative} weights of $E_{10(10)}$ if the contribution
from the lapse function is included. In addition, there are
indications that the relevant representations of $E_{10(10)}$ are
so-called \emph{non-integrable} representations, which are not well
understood.

Given these considerations, it would be desirable to have an
alternative interpretation of the results where one neglects the
overall volume factor (or, in the $E_{10(10)}$-case, the lapse
function) in the analysis of the weight structure.

First, what information does the weight structure contain? Apart
from the overall dilaton factor, the reduction of any higher
derivative term $\sim \mc{R}^p$ will give rise to terms with
$\Pc^{2p}$ (and terms with more derivatives and fewer $\Pc$'s),
where $\Pc$ represents any of the ``building blocks'' $P$, $H$ and
$\partial\phi$ (we suppress all 3-dimensional indices). The
appearance of weights of $\mf{sl}(n+1, \mbb{R})$ (without the
uniform shift from the overall dilaton factor) reflects the fact
that we use fields which are components of the symmetric part of the
left-invariant Maurer--Cartan form $\Pc$ of $\mf{sl}(n+1, \mbb{R})$.
Moreover, the dilaton factor contains information about the number
of such fields. A term $\mc{R}^{l/2}$ will generically give weights
in the weight space of the representation $[l/2,0,\ldots,0,l/2]$ of
$\mf{sl}(n+1, \mbb{R})$, and fill out the positive part of this
weight space.\footnote{We note that the representation structure
encountered here is of the same type as for the lattice of BPS
charges in string theory on $T^{n}$ \cite{ObersPioline}.} This much
is clear from the observation that the overall dilaton factor really
is ``overall''.

The presence of the overall dilaton factor shifts this weight
space uniformly in a negative direction. This shift happens to be
by a vector in the weight lattice of $\mf{sl}(n+1, \mbb{R})$ for
any value of $p$. However, we emphasize that the dilaton exponents
still lie in the weight space of the representation
$[l/2,0,\ldots,0,l/2]$, albeit shifted ``downwards''. From this
point of view, the weight space of the representation with the
shifted highest weight of $[l/2,0,\ldots,0,l/2]$ as highest weight
-- for example, the representation $[2, 0, 1]$ in the case
discussed above -- does not contain all the weights that appear in
the reduced Lagrangian, and therefore does not appear to be relevant.

\subsection{An $SL(n+1, \mbb{Z})$-Invariant Effective Action}

Consider now the fact that it is really the discrete ``U-duality''
group $SL(n+1, \mbb{Z}) \subset SL(n+1, \mbb{R})$ which is expected
to be a symmetry of the complete effective action. Therefore, the
compactified action should be seen as a remnant of the full
U-duality invariant action, arising from a ``large volume
expansion'' of certain automorphic forms.

Schematically, a generic, quartic, scalar term in the action
%(again, restricting to $D=6, n=3$ for convenience)
after compactification of
the Gauss-Bonnet term is of the form
\beq
%\int d^3 x \sqrt{g^{(3)}} e^{-\hat{\vL}_3\cdot \vp} k(\vp)
%F^{[R]}_{IJ}(X),
\int d^3 x \sqrt{|g|} e^{-\hat{\vL}_n\cdot \vp}F(\Pc),
\eeq
where
%$F^{[R]}_{IJ}(X)$
$F(\Pc)$ is a quartic polynomial
%in the scalars $X$ of the theory,
%and $k(\vp)$ is some function of the ``dilatons'' $\vp$.
in the components of the Maurer--Cartan form mentioned above. $F$
will be invariant under $SO(n)$ by construction, but generically not
under $SO(n+1)$.
%This term is not an $SO(4)$ scalar, since $F^{[R]}_{IJ}(X)$ is
%constructed from a nonlinear realisation of the $\mf{sl}(4,
%\mbb{R})$-representation $[2, 0, 2]$, found in Section
%\ref{section:SpecialCaseNoVolumeFactor}. Therefore $F^{[R]}_{IJ}(X)$
%transforms in an $\mf{so}(4)$-representation $R$ which depends on $[2,
%0, 2]$ of $\mf{sl}(4, \mbb{R})$.

To obtain an action which is a scalar under $SO(n+1)$ we must first
``lift'' the result of the compactification to a globally $SL(n+1,
\mbb{Z})$-invariant expression. This can be done by replacing
%the coefficient $k(\vp)$
$e^{-\hat{\vL}_n\cdot \vp}F(\Pc)$
by a suitable automorphic form contracted with four $\Pc$'s:
\beq
\Psi_{I_1\ldots I_8}(X)\Pc^{I_1I_2}\Pc^{I_3I_4}\Pc^{I_5I_6}\Pc^{I_7I_8},
\eqnlab{completion}
\eeq
where the $I$'s are vector indices of $SO(n+1)$. Here, $\Psi(X)$ is
an automorphic form transforming in some representation of
$SO(n+1)$, and is constructed as an Eisenstein series, following,
e.g., refs. \cite{LambertWest2,Aspects}.
%, which transforms in the same
%representation (or, rather, the conjugate representation $\bar{R}$) of
%$SO(4)$ as the polynomial $F^{[R]}_{IJ}(X)$. In addition,
%$\Psi^{IJ}_{[\bar{R}]}(X)$ must reproduce the coefficient $k(\vp)$ as
%its first order term in a ``large volume expansion'',
%\beq
%\Psi^{IJ}_{[\bar{R}]}(X) \sim k(\vp) \quad \text{as}\quad
%\vp\longrightarrow \infty.
%\eeq
We must demand that when the large volume limit, $\hat{\vL}_n\cdot
\vp\rightarrow-\infty$, is imposed, the leading behaviour is
\beq
\Psi_{I_1\ldots
I_8}(X)\Pc^{I_1I_2}\Pc^{I_3I_4}\Pc^{I_5I_6}\Pc^{I_7I_8}\quad
\longrightarrow \quad e^{-\hat{\vL}_n\cdot \vp}F(\Pc).
\eqnlab{largevolumelimit}
\eeq
This limit was taken explicitly in \cite{LambertWest2,Aspects}. This
gives conditions on which irreducible $SO(n+1)$ representations the
automorphic forms transform under (from the tensor structure), as
well as a single condition on the ``weights'' of the automorphic
forms (from the matching of the overall dilaton factor). Automorphic
forms exist for continuous values of the weight (unlike holomorphic
Eisenstein series) above some minimal value derived from convergence
of the Eisenstein series. It was proven in \cite{Aspects} that any
$SO(n)$-covariant tensor structure can be reproduced as the large
volume limit of some automorphic form, and that the weight dictated
by the overall dilaton factor is consistent with the convergence
criterion.

Under the assumption that these arguments are valid, we may conclude
that the representation theoretic structure of the dilaton exponents
in the polynomial $F$ should be analyzed without inclusion of the
volume factor $e^{-\hat{\vL}_n\cdot \vp}$, and hence, for the
Gauss-Bonnet term ($l=4$), it is the $[2, 0, \dots, 0,
2]$-representation which is the relevant one (in the sense above,
that we are dealing with products of four Maurer--Cartan forms), and
\emph{not} the representation $[2, 0, \dots, 0, 1]$. Another indication for why the
representation with highest weight $2\hat{\vL}_1+\hat{\vL}_n$ cannot
be the relevant one is that it is not contained in the tensor
product of the adjoint representation $[1, 0, \dots, 0, 1]$ of
$\mf{sl}(n+1, \mbb{R})$ with itself.

The present point of view also suggest a possible explanation for
the discrepancy of the weight multiplicities observed in the
previous section. In the complete $SL(n+1, \mbb{Z})$-invariant
four-derivative effective action the multiplicities of the weights
in the $[2, 0, \dots, 0, 2]$-representation necessarily match
because the action is constructed directly from the $\mf{sl}(n+1,
\mbb{R})$-valued building block $\mc{P}$. When taking the large
volume limit, \Eqnref{largevolumelimit}, a lot of information is
lost (see, e.g., \cite{Aspects}) and it is therefore natural that
the result of the compactification does not display the correct
weight multiplicities. Thus, it is only after taking the
non-perturbative completion, \Eqnref{completion}, that we can expect
to reproduce correctly the weight multiplicities of the
representation $[2, 0, \dots, 0, 2]$.

\subsection{Algebraic Constraints on Curvature Corrections}

Our results have additional implications for the interpretations of
the weight structure laid forward in \cite{LambertWest1}. In the
analysis of the compactification of eleven-dimensional supergravity
to three dimensions these authors include the volume factor when
investigating the weight structure of $E_{8(8)}$. This implies that
an arbitrary weight for the $l/2$:th order correction terms contains
a factor of $(\f{1}{3}-\f{l}{6})\hat{\vL}_8$. In our example above
this precisely corresponds to the volume factor $\hat{\vL}_n$.
Including this factor and demanding that all dilaton exponents
should be on the weight lattice of $E_{8(8)}$ gives the constraint
\beq
\f{1}{3}-\f{l}{6}\in \mbb{Z}\quad \LongleftrightarrowÊ\quad l=6k+2,
\quad (k=0, 1, 2, \dots ).
\eqnlab{powerrestriction}
\eeq
This implies that these can only be on the weight lattice of
$E_{8(8)}$ if the orders of the curvature correction are the
celebrated powers $\f{l}{2}=3k+1, \hs k=0, 1, 2, \dots$. However, if
our interpretation is correct, the volume factor should be left
outside of the representation structure and so this argument about
the restrictions on $l$ does not seem to be applicable from a purely
mathematical point of view, since also intermediate values can be
reproduced by automorphic forms with some (continuous) weight.\footnote{The fact that $E_{8(8)}$-invariant terms which do not arise from the compactification of $\mc{R}^{3k+1}$ curvature corrections can exist in $D=3$ follows also from the work of \cite{LambertWest2}, which however
emphasizes a different role of the dilaton pre-factors
compared to the one suggested here. We thank the authors of \cite{LambertWest2} for
correspondence on this issue.}
%It is possible that perhaps the
%constraint in \Eqnref{powerrestriction} could be enforced from a
%different point of view, by analyzing the automorphic form which
%reproduces the exponential power $(\f{1}{3}-\f{l}{6})\hat{\vL}_8$ in
%its large volume expansion.

This, of course, does not mean that the result itself is incorrect
(it is well known, e.g., that the first higher-derivative correction
allowed by supersymmetry is of order $\mc{R}^4$, as is the first
correction obtained by superstring calculations), only that the
arguments used in order to reach it have to be refined. In order to
obtain the result in the present context, one would need information
restricting the weights of the automorphic forms that may enter to
some discrete values. Real automorphic forms defined by Eisenstein
series, unlike the holomorphic ones of $SL(2, \mbb{R})$ (or $Sp(2n)$
in general), are defined for continuous values of the weight,
bounded from below only by the convergence of the series. When
one-loop calculations in eleven-dimensional supergravity have been
used to derive automorphic forms occurring in $d=9$
\cite{GreenGutperle}, it is clear how well-defined values of the
weights arise. The corresponding picture for compactification to
lower dimensions is less clear, due to the presence of membrane and
5-brane instantons \cite{ObersPioline,KiritsisPioline}, but there is
no doubt a corresponding mechanism at play, although we lack enough
insight into the microscopic degrees of freedom to make a clear
statement about it.

We suspect that a reasoning along similar lines may be used for the
case of $E_{10(10)}$, and that it may again lead to the conclusion
that the shifted highest weight should not be interpreted as the
highest weight of a new (non-integrable) representation. Instead, it
may be possible to deal with automorphic forms transforming in some
integrable representations of the maximal compact subgroup of
$E_{10(10)}$.

\vspace{1cm}

\section*{Acknowledgements}
We thank Jarah Evslin, Marc Henneaux, Axel Kleinschmidt, Stanislav
Kuperstein, Jakob Palmkvist and Christoffer Petersson for
discussions. We would also like to thank Neil Lambert and Peter West for correspondence. 

D.P. is supported in part by IISN-Belgium (convention 4.4505.86), by
the Belgian National Lottery, by the European Commission FP6 RTN
programme MRTN-CT-2004-005104 for which he is associated with VUB,
and by the Belgian Federal Science Policy Office through the
Interuniversity Attraction Pole P5/27.
\newpage
\begin{appendix}

\section{Squared Curvature terms}

\label{appendix}

Here we present all the detailed computations of the
compactification.

\subsection{Weyl-Rescaling}

Weyl-rescaling the $D$-dimensional metric by a factor $e^{2
\varphi}$:
\beq
\hat{g}_{MN} = e^{2 \varphi} \tilde{g}_{MN},
\eeq
yields the rescaled Riemann tensor
\beqa
{}Ê\hat{R}_{ABCD} &= &e^{-2\varphi}\big[\tilde{R}_{ABCD} - 2 \big(
\eta_{[A|C|}\tilde{\nabla}_{B]}\tilde{\partial}_D\varphi -
\eta_{[A|D|}\tilde{\nabla}_{B]}\tilde{\partial}_C\varphi \big)
\nn \\
{}Ê& & + 2\big(
\eta_{[A|C|}\tilde{\partial}_{B]}\varphi \tilde{\partial}_D\varphi -
\eta_{[A|D|}\tilde{\partial}_{B]}\varphi \tilde{\partial}_C\varphi
\big) - 2\eta_{[A|C|}\eta_{B]D}(\tilde{\partial}\varphi)^2
\big],
\eeqa
Ricci tensor
\beq
\hat{R}_{AB} = e^{-2\varphi}\big[ \tilde{R}_{AB} -
\eta_{AB}\tilde{\square}\varphi -
(D-2)\tilde{\nabla}_A\tilde{\partial}_B\varphi +
(D-2)\tilde{\partial}_A\varphi\tilde{\partial}_B\varphi -
(D-2)\eta_{AB}(\tilde{\partial}\varphi)^2\big],
\eeq
and curvature scalar
\beq
\hat{R} = e^{-2\varphi}\big[ \tilde{R} -
(D-1)(D-2)(\tilde{\partial}\varphi)^2 - 2(D-1)\tilde{\square}\varphi
\big].
\eeq
Squaring the curvature terms we find
\beqa
{}Ê(\hat{R}_{ABCD})^2 &= &e^{-4\varphi}\big[ (\tilde{R}_{ABCD})^2 +
8\big(\tilde{R}_{AB}-\frac{1}{2}\eta_{AB}\tilde{R}\big)
\tilde{\partial}^A\varphi\tilde{\partial}^B\varphi -
8\tilde{R}_{AB}\tilde{\nabla}^A\tilde{\partial}^B\varphi +
4(\tilde{\square}\varphi)^2
\nn \\
{}Ê& & +
4(D-2)(\tilde{\nabla}_A\tilde{\partial}_B\varphi)
(\tilde{\nabla}^A\tilde{\partial}^B\varphi) +
8(D-2)(\tilde{\partial}\varphi)^2\tilde{\square}\varphi
\nn \\
{}Ê& &-
8(D-2)\tilde{\partial}^A\varphi\tilde{\partial}^B\varphi
\tilde{\nabla}_A\tilde{\partial}_B\varphi
 + 2(D-1)(D-2)(\tilde{\partial}\varphi)^2
(\tilde{\partial}\varphi)^2 \big],
\eeqa
\beqa
{}Ê(\hat{R}_{AB})^2 &= & e^{-4 \varphi} \big[ (\tilde{R}_{AB})^2 -
2\tilde{R}\tilde{\square}\varphi -
2(D-2)\tilde{R}_{AB}\tilde{\nabla}^A\tilde{\partial}^B\varphi +
(3D-4)(\tilde{\square}\varphi)^2
\nn \\
{}Ê& &   +
2(D-2)\big(\tilde{R}_{AB}-\eta_{AB}\tilde{R}\big)
(\tilde{\partial}^A\varphi)(\tilde{\partial}^B\varphi) +
(D-2)^2(\tilde{\nabla}_A\tilde{\partial}_B\varphi)
(\tilde{\nabla}^A\tilde{\partial}^B\varphi)
\nn \\
{}Ê& &
 + (D-1)(D-2)^2(\tilde{\partial}\varphi)^4 +
2(D-2)(2D-3)(\tilde{\square}\varphi)(\tilde{\partial}\varphi)^2
\nn \\
{}Ê& &  -
2(D-2)^2(\tilde{\nabla}_A\tilde{\partial}_B\varphi)
(\tilde{\partial}^A\varphi)(\tilde{\partial}^B\varphi)\big],
\eeqa
\beqa
{}Ê\hat{R}^2 &=& e^{-4\varphi}\big[ \tilde{R}^2 -
4(D-1)\tilde{R}\tilde{\square}\varphi -
2(D-1)(D-2)\tilde{R}(\tilde{\partial}\varphi)^2 +
4(D-1)^2(\tilde{\square}\varphi)^2
\nn \\
{}Ê& & +
4(D-1)^2(D-2)(\tilde{\square}\varphi)(\tilde{\partial}\varphi)^2 +
(D-1)^2(D-2)^2(\tilde{\partial}\varphi)^4 \big].
\eeqa

Combining these, the Gauss-Bonnet combination can be written as
\beqa
{}Ê\hat{R}_{\mathrm{GB}}^2 &= &e^{-4\varphi}\big\{
\tilde{R}_{\mathrm{GB}}^2 + (D-3)\big[ 8\big(\tilde{R}_{AB} -
\frac{1}{2}\eta_{AB}\tilde{R}\big)\tilde{\nabla}^A\tilde{\partial}^B
\varphi - 8\tilde{R}_{AB}\tilde{\partial}^A\varphi
\tilde{\partial}^B\varphi
 \nn \\
{}Ê& & - 2(D-4)\tilde{R}(\tilde{\partial}\varphi)^2 +
4(D-2)(D-3)(\tilde{\partial}\varphi)^2\tilde{\square}\varphi +
8(D-2)(\tilde{\nabla}_A\tilde{\partial}_B\varphi)
\tilde{\partial}^A\varphi\tilde{\partial}^B\varphi \nn \\
{}Ê& &   + 4(D-2)[(\tilde{\square}\varphi)^2 -
(\tilde{\nabla}_A\tilde{\partial}_B\varphi)
(\tilde{\nabla}^A\tilde{\partial}^B\varphi)] +
(D-1)(D-2)(D-4)(\tilde{\partial}\varphi)^4 \big] \big\}.
\nn \\
\eeqa

The Gauss-Bonnet Lagrangian, including the measure $\hat{e} =
e^{D\varphi}\tilde{e}$, can now be conveniently grouped in terms of
equations of motion and total derivatives. This is achieved using
integrations by parts, where no explicit appearance of
$\tilde{\nabla}_{(A}\tilde{\partial}_{B)}\varphi$ is required. The
resulting Lagrangian is
\beqa
{}Ê\mathcal{L}_{\mathrm{GB}} &= &\tilde{e} e^{(D-4)\varphi} \big\{
\tilde{R}_{\mathrm{GB}}^2 - (D-3)(D-4)\big[
2(D-2)(\tilde{\partial}\varphi)^2\tilde{\square}\varphi
+ (D-2)(D-3)(\tilde{\partial}\varphi)^4 \big . \nn \\
{}Ê& &  + 4 \big(\tilde{R}_{AB} -
\frac{1}{2}\eta_{AB}\tilde{R}\big)
(\tilde{\partial}^A\varphi)(\tilde{\partial}^B\varphi)\big]\big\}\nn \\
{}Ê& & + 2(D-3)\tilde{e}\tilde{\nabla}_A\big\{
e^{(D-4)\varphi}\big[(D-2)^2(\tilde{\partial}
\varphi)^2\tilde{\partial}^A\varphi + 2(D-2)(\tilde{\square}
\varphi)\tilde{\partial}^A\varphi \nn  \\
{}Ê& & - (D-2)\tilde{\partial}^A(\tilde{\partial} \varphi)^2 +
4\big(\tilde{R}^{AB} - \frac{1}{2}\eta^{AB}\tilde{R}\big)
\tilde{\partial}_B\varphi\big]\big\}.
\eqnlab{eqn:GBweyl2}
\eeqa
Notice that the total derivative terms in this expression will
remain total derivatives after the compactification as well.

\subsection{Compactification}

In compactification of gravity from $D$ to $d=(D-n)$ dimensions
the vielbein one-form is given by
\beq
\tilde{e}^A = (\tilde{e}^\alpha,\tilde{e}^a) = \big(e^\alpha,
[dx^m+\mathcal{A}_{(1)}^m]\tvb{m}{a}\big), \label{eqn:tvbform}
\eeq
with the determinant denoted by $\tilde{e}=e\tvbint$. Dropping all
dependence on the torus coordinates, i.e.,
$\tilde{d}=d=dx^\mu\partial_\mu$, the compactified spin connection
one-form is found to be
\beqa
{}Ê\tilde{\omega}^{\alpha}_{\phantom{\alpha}\beta} &=&
\omega^{\alpha}_{\phantom{\alpha}\beta} - \frac{1}{2}\tilde{e}^c
\tF_{c\phantom{\alpha}\beta}^{\phantom{c}\alpha},
\nn \\
{}Ê\tilde{\omega}^{\alpha}_{\phantom{\alpha}b}& =& \frac{1}{2}
e^\gamma\tF_{b\gamma}^{\phantom{b\gamma}\alpha} -
\tilde{e}^c\tP^{\alpha}_{\phantom{\alpha}cb},
\nn \\
{}Ê\tilde{\omega}^{a}_{\phantom{a}b} &= & e^\gamma Q_{\gamma
\phantom{a}b}^{\phantom{\gamma}a},
\eeqa
where $\tP_\alpha^{\phantom{\alpha} bc} =
\tilde{e}^{m(b}\partial_\alpha\tvb{m}{c)}$,
$Q_\alpha^{\phantom{\alpha} bc} =
\tilde{e}^{m[b}\partial_\alpha\tvb{m}{c]}$ and
$\tF^a_{\phantom{a}\beta\gamma} = 2\tvb{m}{a}
e^{\mu}_{\phantom{\mu}[\beta}e^{\nu}_{\phantom{\nu}\gamma]}
\partial_{\mu}^{}\mathcal{A}_{\nu}^m$.

Using the spin connection it is now straightforward to compute the
compactified Riemann tensor
\beqa
{}Ê\tilde{R}_{\alpha \beta \gamma \delta} &=&
R_{\alpha\beta\gamma\delta} - \frac{1}{2}\big(\tF^{}_{c\alpha
[\gamma}\tF^c_{\phantom{c}|\beta|\delta]} + \tF_{c\alpha
\beta}\tF^c_{\phantom{c}\gamma\delta}\big),
\nn \\
{}Ê\tilde{R}_{\alpha\beta\gamma d} &=&
D_{[\alpha}\tF_{|d|\beta]\gamma} -
\tF^c_{\phantom{c}\alpha\beta}\tP_{\gamma cd},
\nn \\
{}Ê\tilde{R}_{\alpha\beta cd} &=&
\frac{1}{2}\tF_{[c|\alpha|}^{\phantom{c
|\alpha|}\gamma}\tF_{d]\gamma\beta}^{\phantom{\gamma}} -
2\tP^{\phantom{\alpha}e}_{\alpha\phantom{e}[c}\tP_{|\beta| d]e},
\nn \\
{}Ê\tilde{R}_{\alpha b \gamma d} &=& -D_\alpha\tP_{\gamma bd} -
\tP_{\alpha b}^{\phantom{\alpha b}e}\tP_{\gamma de} +
\frac{1}{4}\tF_{b\gamma\epsilon}\tF_{d\alpha}^{\phantom{d
\alpha}\epsilon},
\nn \\
{}Ê\tilde{R}_{ab\gamma d} &=&
\tF_{[a|\gamma\epsilon|}\tP^\epsilon_{\phantom{\epsilon}b]d},
\nn \\
{}Ê\tilde{R}_{abcd} &= &- 2\tP_{\epsilon a[c}
\tP^\epsilon_{\phantom{\epsilon}|b|d]},
\eeqa
which contracted yields the Ricci tensor
\beqa
{}Ê\tilde{R}_{\alpha\beta}&=& R_{\alpha\beta} -
\frac{1}{2}\tF_{c\epsilon\alpha}\tF^{c\epsilon}_{\phantom{c\epsilon}\beta}
- \tP_{\alpha cd}\tP^{\phantom{\beta}cd}_\beta - \tr(D_\alpha
\tP_\beta),
\nn \\
{}Ê\tilde{R}_{\alpha b} &= & \frac{1}{2}\big(D_\epsilon
\tF_{b\alpha}^{\phantom{b\alpha}\epsilon} +
\tF_{c\alpha\delta}\tP^{\delta c}_{\phantom{\delta c}b}
+ \tF_{b\alpha\epsilon}\trtP^\epsilon\big),
\nn \\
{}Ê\tilde{R}_{ab} &= &- D^\epsilon \tP_{\epsilon ab} - \tP_{\epsilon
ab}\trtP^\epsilon +
\frac{1}{4}\tF_{a\gamma\delta}\tF_b^{\phantom{b}\gamma\delta},
\eeqa
and the curvature scalar
\beq
\tilde{R} = R - \frac{1}{4}\tF^2 - \tP^2 - (\trtP)^2 -
2\tr(D_\epsilon\tP^\epsilon).
\eeq
The trace is always taken over the internal indices, also
$\tF^2\equiv\tF_{a\beta\gamma}\tF^{a\beta\gamma}$ and $\tP^2\equiv
\tP_{\alpha bc}\tP^{\alpha bc}$. The covariant derivative $D$ is
defined as $D \equiv \nabla + Q \equiv \partial + \omega + Q$, where
$\omega_{\alpha\beta\gamma}$ is the spacetime spin connection and
$Q_{\alpha bc}$ can be thought of as a gauge connection for the
$SO(n)$-symmetry.

Squaring the curvature tensor components we find:
\beqa
{}Ê(\tilde{R}_{\alpha\beta\gamma\delta})^2 &=&
R_{\alpha\beta\gamma\delta}R^{\alpha\beta\gamma\delta} -
\frac{3}{2}R_{\alpha\beta\gamma\delta}
\tF_e^{\phantom{e}\alpha\beta}\tF^{e\gamma\delta} +
\frac{3}{8}\tFFFFa
\nn \\
{}Ê& & + \frac{3}{8}\tFFFFc,
\nn \\
{}Ê(\tilde{R}_{\alpha\beta\gamma d})^2 &=&
(D_{[\alpha}\tF_{|d|\beta]\gamma})(D^\alpha\tF^{d\beta\gamma}) -
2(D_\alpha\tF_{d\beta\gamma})\tF^{c\alpha\beta}\tP^{\gamma
d}_{\phantom{\gamma d}c} + \tF^a_{\phantom{a}\gamma\delta}
\tP_{\epsilon ab} \tP^{\epsilon bc}\tF_c^{\phantom{c}\gamma\delta},
\nn \\
{}Ê(\tilde{R}_{\alpha\beta cd})^2 &=&
2\tr(\tP_\alpha\tP^\alpha\tP_\beta\tP^\beta) -
2\tr(\tP_\alpha\tP_\beta\tP^\alpha\tP^\beta) + \frac{1}{8}\tFFFFb
\nn \\
{}Ê& & - \frac{1}{8}\tFFFFc - \tF^c_{\phantom{c}\beta\gamma}
\tP_{\delta ce} \tP^{\gamma ed}\tF_d^{\phantom{d}\beta\delta} +
\tF^c_{\phantom{c}\beta\gamma} \tP^\gamma_{\phantom{\gamma} ce}
\tP_\delta^{\phantom{\delta}
ed}\tF_d^{\phantom{d}\beta\delta},
\nn \\
{}Ê(\tilde{R}_{\alpha b \gamma d})^2 &= &(D_\alpha \tP_{\gamma bd})
(D^\alpha \tP^{\gamma bd}) + 2(D_\alpha \tP_{\gamma
bd})\tP^{\alpha be}\tP^{\gamma d}_{\phantom{\gamma d}e} -
\frac{1}{2}(D_\alpha \tP_{\gamma bd})\tF^{b\alpha\epsilon}
\tF^{d\gamma}_{\phantom{d\gamma}\epsilon}
\nn \\
{}Ê& & +
\tr(\tP_\alpha\tP^\alpha\tP_\gamma\tP^\gamma) +
\frac{1}{16}\tFFFFc - \frac{1}{2}\tF^b_{\phantom{b}\beta\gamma}
\tP_{\delta be}
\tP^{\gamma ed}\tF_d^{\phantom{d}\beta\delta},
\nn \\
{}Ê(\tilde{R}_{ab\gamma d})^2 &=& \frac{1}{2}\big[
\tF_{c\phantom{\alpha}\delta}^{\phantom{c}\alpha}\tF^{c\beta\delta}
\tr(\tP_\alpha\tP_\beta) - \tF^a_{\phantom{a}\beta\gamma}
\tP_{\delta ad} \tP^{\gamma db}\tF_b^{\phantom{b}\beta\delta} \big],
\nn \\
{}Ê(\tilde{R}_{abcd})^2 &=& 2\tr(\tP_\alpha\tP_\beta)
\tr(\tP^\alpha\tP^\beta) -
2\tr(\tP_\alpha\tP_\beta\tP^\alpha\tP^\beta).
\eeqa
The compactified Ricci tensor and curvature scalar squared are
\beqa
{}Ê(\tilde{R}_{\alpha\beta})^2 &= &R_{\alpha\beta}R^{\alpha\beta} -
R_{\alpha\beta}\tF_{c\delta}^{\phantom{c\delta}\alpha}
\tF^{c\delta\beta} - 2R_{\alpha\beta} \tr(\tP^\alpha\tP^\beta) -
2R_{\alpha\beta}\tr(D^\alpha\tP^\beta)
\nn \\
{}Ê& & +
\tr(D_\alpha\tP_\beta)\tr(D^\alpha\tP^\beta) +
2\tr(D_\alpha\tP_\beta)\tr(\tP^\alpha\tP^\beta) +
\tr(\tP_\alpha\tP_\beta)\tr(\tP^\alpha\tP^\beta)
\nn \\
{}Ê& &  + \tr(D_\alpha\tP_\beta)
\tF_{c\delta}^{\phantom{c\delta}\alpha}\tF^{c\delta\beta} +
\tr(\tP_\alpha\tP_\beta)
\tF_{c\delta}^{\phantom{c\delta}\alpha}\tF^{c\delta\beta} +
\frac{1}{4}\tFFFFb,
\nn \\
{}Ê(\tilde{R}_{\alpha b})^2 &=& \frac{1}{4} \big[(D_\gamma
\tF_{b\alpha}^{\phantom{b\alpha}\gamma})(D_\delta
\tF^{b\alpha\delta}) + 2(D_\gamma
\tF_{b\alpha}^{\phantom{b\alpha}\gamma})\tF^{b\alpha\beta}
\trtP_\beta + 2(D_\gamma \tF_{c\alpha}^{\phantom{c\alpha}\gamma})
\tP_\beta^{\phantom{\beta}cd}\tF_d^{\phantom{d}\alpha\beta}
\nn \\
{}Ê& &   + \tF^c_{\phantom{c}\beta\gamma}
\tP^\gamma_{\phantom{\gamma} ce} \tP_\delta^{\phantom{\delta}
ed}\tF_d^{\phantom{d}\beta\delta} + 2\tF_{b\alpha\gamma}
\tP_\beta^{\phantom{\beta}bc}\tF_c^{\phantom{c}\alpha\beta}\trtP^\gamma
+ \tF_{b\alpha\gamma}\tF^{b\alpha}_{\phantom{b\alpha}\delta}
(\trtP^\gamma)(\trtP^\delta)\big],
\nn \\
{}Ê(\tilde{R}_{ab})^2 &=& \tr(D_\alpha\tP^\alpha D_\beta\tP^\beta) +
2(D_\alpha\tP^{\alpha bc})\tP_{\beta bc}\trtP^{\beta} -
\frac{1}{2}(D_\alpha\tP^{\alpha bc})\tF_{b\gamma\delta}
\tF_c^{\phantom{c}\gamma\delta}
\nn \\
{}Ê& &  +
\tr(\tP_\alpha\tP_\beta)(\trtP^\alpha)(\trtP^\beta) -
\frac{1}{2}\tF_{b\alpha\beta}
\tP_\gamma^{\phantom{\gamma}bc}\tF_c^{\phantom{c}\alpha\beta}\trtP^\gamma
+ \frac{1}{16}\tF_{c\alpha\beta}\tF_d^{\phantom{d}\alpha\beta}
\tF^c_{\phantom{c}\gamma\delta}\tF^{d\gamma\delta},
\nn \\
\eeqa
and
\beqa
{}Ê\tilde{R}^2 &=& R^2 - \frac{1}{2}R\tF^2 - 4R\tr(D_\alpha\tP^\alpha) -
2R\tP^2 - 2R(\trtP)^2 + \frac{1}{16}(\tF^2)^2 +
\tF^2\tr(D_\alpha\tP^\alpha)
\nn \\
{}Ê& &+ \frac{1}{2}\tF^2\tP^2 +
\frac{1}{2}\tF^2(\trtP)^2 + 4[\tr(D_\alpha\tP^\alpha)]^2 +
4\tr(D_\alpha\tP^\alpha)\tP^2 + 4\tr(D_\alpha\tP^\alpha)(\trtP)^2
\nn \\
{}Ê& & + (\tP^2)^2 + 2\tP^2(\trtP)^2 + ((\trtP)^2)^2.
\eeqa

Choosing a basis where all explicit derivatives appearing are either
divergences or total derivatives, we can rewrite three of the
quadratic curvature components as
\beqa
{}Ê (\tilde{R}_{\alpha\beta\gamma d})^2 &=& \frac{1}{2} \big[
R_{\alpha\beta\gamma\delta}\tF_e^{\phantom{e}\alpha\gamma}
\tF^{e\beta\delta} -
R_{\alpha\beta}\tF_{e\gamma}^{\phantom{e\gamma}\alpha}
\tF^{e\gamma\beta} + (D_\gamma
\tF_{b\alpha}^{\phantom{b\alpha}\gamma})(D_\delta
\tF^{b\alpha\delta})
\nn \\
{}Ê & & - \tF_{c\alpha\beta}(D_\gamma\tP^{\gamma
cd})\tF_d^{\phantom{d}\alpha\beta} + \tFPPFa - \tFPPFb
\nn \\
{}Ê & & + 3\tFPPFc \big] + \frac{1}{2}\nabla_\alpha \big[ (D_\gamma
\tF_{e\beta}^{\phantom{e\beta}\alpha})\tF^{e\beta\gamma} - (D_\gamma
\tF_{e\beta}^{\phantom{e\beta}\gamma})\tF^{e\beta\alpha}
\nn \\
{}Ê & & + \tF_{c\beta\gamma}\tP^{\alpha
cd}\tF_d^{\phantom{d}\beta\gamma}\big],
\nn \\
{}Ê (\tilde{R}_{\alpha b\gamma d})^2 &=& \frac{1}{2}(D_\alpha
\tF_{c\beta}^{\phantom{c\beta}\alpha})
\tP_\gamma^{\phantom{\gamma}cd}\tF_d^{\phantom{d}\beta\gamma} -
\frac{1}{8}\tFDPF + \tr[(D_\alpha\tP^\alpha)(D_\beta\tP^\beta)]
\nn \\
{}Ê & & - \tr[(D_\alpha\tP^\alpha)\tP_\beta\tP^\beta] -
R_{\alpha\beta}\tr(\tP^\alpha\tP^\beta) + \frac{1}{16}\tFFFFb
\nn \\
{}Ê & & - \frac{1}{4}\tFPPFa +
2\tr(\tP_\alpha\tP_\beta\tP^\alpha\tP^\beta) -
\tr(\tP_\alpha\tP^\alpha\tP_\beta\tP^\beta) + \nabla_\alpha\big[
\tr(\tP^\alpha\tP_\beta\tP^\beta)
\nn \\
{}Ê & & + \tr(\tP_\beta D^\beta\tP^\alpha - \tP^\alpha
D^\beta\tP_\beta) + \frac{1}{8}\tFPFa - \frac{1}{2}\tFPFb \big], \nn
\\
{}Ê (\tilde{R}_{\alpha\beta})^2 &=& R_{\alpha\beta}R^{\alpha\beta} -
R_{\alpha\beta}\tF_{c\delta}^{\phantom{c\delta}\alpha}\tF^{c\delta\beta}
- 2R_{\alpha\beta}\tr(\tP^\alpha \tP^\beta) - R_{\alpha\beta}
\trtP^\alpha\trtP^\beta - R\tr(D_\alpha\tP^\alpha) \nn \\
{}Ê & & - (D_\alpha \tF_{c\delta}^{\phantom{c\delta}\alpha})
\tF^{c\delta\beta}\trtP_\beta - 2\tr[(D_\alpha\tP^\alpha)
\tP^\beta]\trtP_\beta + \tr(D_\alpha\tP^\alpha)
\tr(D_\beta\tP^\beta) \nn \\
{}Ê & & + \frac{1}{4}\tr(D_\alpha\tP^\alpha)\tF^2 +
\tr(D_\alpha\tP^\alpha) \tr(\tP_\beta\tP^\beta) + \frac{1}{4}\tFFFFb
\nn \\
{}Ê & & + \frac{1}{2}\tFPFa \trtP_\alpha - \tFPFb \trtP_\alpha +
\tF_{c\delta\alpha} \tF^{c\delta}_{\phantom{c\delta}\beta}
\tr(\tP^\alpha \tP^\beta) \nn \\
{}Ê & & + \tr(\tP_\alpha\tP_\beta)\tr(\tP^\alpha\tP^\beta) +
\nabla_\alpha \big[\tr(D^\alpha \tP_\beta)\trtP^\beta - \tr(D^\beta
\tP_\beta)\trtP^\alpha - \frac{1}{4}\tF^2\trtP^\alpha \nn \\
{}Ê & & - 2(R^{\alpha\beta} - \frac{1}{2}\eta^{\alpha\beta}R)
\trtP_\beta + \tF^{c\delta\alpha}
\tF_{c\delta}^{\phantom{c\delta}\beta}\trtP_\beta + 2\tr(\tP^\alpha
\tP^\beta)\trtP_\beta - \tP^2\trtP^\alpha \big].
\eeqa
This choice of basis is only possible for the curvature terms
squared; for general powers of the Riemann tensor no such basis
exists. The square of the uncompactified curvature terms can now be
written as a sum of the quadratic compactified curvature components
\beqa
{}Ê(\tilde{R}_{ABCD})^2 &=& \tilde{R}_{\alpha\beta\gamma\delta}
\tilde{R}^{\alpha\beta\gamma\delta} +
4\tilde{R}_{\alpha\beta\gamma d} \tilde{R}^{\alpha\beta\gamma d} +
2\tilde{R}_{\alpha\beta cd} \tilde{R}^{\alpha\beta cd} +
4\tilde{R}_{\alpha b \gamma d}\tilde{R}^{\alpha b \gamma d}
\nn \\
{}Ê& & + 4\tilde{R}_{ab \gamma d} \tilde{R}^{ab \gamma d} +
\tilde{R}_{abcd}\tilde{R}^{abcd},
\nn \\
{}Ê(\tilde{R}_{AB})^2 &=& \tilde{R}_{\alpha\beta}
\tilde{R}^{\alpha\beta} + 2\tilde{R}_{\alpha b}\tilde{R}^{\alpha
b} + \tilde{R}_{ab}\tilde{R}^{ab}.
\eeqa
Since the total volume measure is $\hat{e} = e^{D\varphi}e\tvbint$,
the factor $e^{D\varphi}\tvbint$ has to be moved inside the total
derivatives using integration by parts. The Riemann tensor squared
will then be given by
\beqa
{}Ê& & \quad \hat{e}e^{-4\varphi}\big(\tilde{R}_{ABCD}\big)^2
\nn \\
{}Ê& &=
ee^{(D-4)\varphi}\tvbint \big\{ R_{\alpha\beta\gamma\delta}
R^{\alpha\beta\gamma\delta} - \frac{1}{2}
R_{\alpha\beta\gamma\delta} \tF_e^{\phantom{e}\alpha\beta}
\tF^{e\gamma\delta} - 2R_{\alpha\beta}\big[
\tF_{c\delta}^{\phantom{c\delta}\alpha}\tF^{c\delta\beta} +
2\tr(\tP^\alpha\tP^\beta) \big]
\nn \\
{} & & + 2
D_\alpha\tF_{c\delta}^{\phantom{c\delta}\alpha} \big[D_\beta
\tF^{c\delta\beta} + \tP_\beta^{\phantom{\beta}cd}
\tF_d^{\phantom{d}\delta\beta} + \trtP_\beta\tF^{c\delta\beta} +
(D-4)\partial_\beta\varphi\tF^{c\delta\beta} \big] + 4
\tr(D_\alpha\tP^\alpha D_\beta\tP^\beta)
\nn \\
{} & &  - 4
\tr[(D_\alpha\tP^\alpha)\tP_\beta\tP^\beta] - \frac{5}{2} \tFDPF +
4\tr(\tP_\alpha D_\beta\tP^\beta)\big[\trtP^\alpha +
(D-4)\partial^\alpha\varphi\big]
\nn \\
{}Ê& &  +
\frac{1}{2}\tr(D_\alpha\tP^\alpha) \big[ \tF^2 + 4\tP^2 \big] +
\frac{3}{8}\tFFFFa + \frac{1}{2}\tFFFFb
\nn \\
{}Ê& & +
\frac{1}{8}\tFFFFc + 2\tFPPFc + \tFPPFa + 2\tr(\tP_\alpha\tP_\beta
\tP^\alpha\tP^\beta)
\nn \\
{} & & + 2\tr(\tP_\alpha\tP_\beta)
\tF_{c\delta}^{\phantom{c\delta}\alpha} \tF^{c\delta\beta} -
\frac{3}{2} \tFPFa\big[ \trtP_\alpha + (D-4)\partial_\alpha\varphi
\big] + 2\tr(\tP_\alpha\tP_\beta)\tr(\tP^\alpha\tP^\beta)
\nn \\
{}Ê& & + \frac{1}{2}\big(\tF^2 + 4\tP^2\big)\big[(\trtP)^2 +
2(D-4)\trtP_\alpha\partial^\alpha\varphi +
(D-4)^2(\partial\varphi)^2 + (D-4)\square\varphi\big]
\nn \\
{} & &
 - 4\tr(\tP_\alpha\tP^\alpha\tP_\beta)\big[ \trtP^\beta +
(D-4)\partial^\beta\varphi \big] \big\} + e\nabla_\alpha \big\{
e^{(D-4)\varphi}\tvbint \big[ - 2(D_\beta
\tF_{c\delta}^{\phantom{c\delta}\beta}) \tF^{c\delta\alpha}
\nn \\
{}Ê& &  - 4\tr(\tP^\alpha D_\beta\tP^\beta) +
\frac{3}{2} \tFPFa + 4\tr(\tP^\alpha\tP_\beta\tP^\beta) -
\big[\trtP^\alpha + (D-4)\partial^\alpha\varphi\big](\tF^2 +
4\tP^2) \big]
\nn \\
{}Ê& &   + D^\alpha\big[
e^{(D-4)\varphi}\tvbint (\frac{1}{2}\tF^2 + 2\tP^2) \big] \big\}
\eqnlab{RiemannSquared}
\eeqa
and the Ricci tensor squared is given by
\beqa
{} & & \quad \hat{e}e^{-4\varphi}\big(\tilde{R}_{AB}\big)^2
\nn \\
{} & &= ee^{(D-4)\varphi}\tvbint \big\{ R_{\alpha\beta} \big[
R^{\alpha\beta} - \tF_{c\delta}^{\phantom{c\delta}\alpha}
\tF^{c\delta\beta} - 2\tr(\tP^\alpha\tP^\beta) + \trtP^\alpha
\trtP^\beta + 2(D-4)\trtP^\alpha \partial^\beta\varphi \big]
\nn \\
{} & & - R\big[ \tr(D_\alpha\tP^\alpha) + (\trtP)^2 +
(D-4)\trtP_\alpha \partial^\alpha\varphi \big] + (D_\alpha
\tF^{\phantom{a\beta}\alpha}_{b\gamma})\big[\frac{1}{2} D_\beta
\tF^{b\gamma\beta} + \tP_\delta^{\phantom{\beta}bc}
\tF_c^{\phantom{c}\gamma\delta} \big]
\nn \\
{}Ê& & + (D_\alpha
\tP^\alpha_{\phantom{\alpha}bc})\big[D_\beta \tP^{\beta bc} -
\frac{1}{2}\tF^b_{\phantom{b}\gamma\delta}
\tF^{c\gamma\delta}\big] + \frac{1}{4}\tFFFFb +
\frac{1}{16}\tFFFFa
\nn \\
{}Ê& & + \tr(D_\alpha \tP^\alpha) \big[
\tr(D_\beta\tP^\beta) + \frac{1}{4}\tF^2 + \tP^2 +
\frac{3}{2}(\trtP)^2 + (D-4)\trtP_\beta\partial^\beta\varphi \big]
\nn \\
{}Ê& & + \frac{1}{2}\tFPPFb + \tF_{c\delta\alpha}
\tF^{c\delta}_{\phantom{c\delta}\beta}\big[ \tr(\tP^\alpha
\tP^\beta) - \frac{1}{2}\trtP^\alpha \trtP^\beta - (D-4)\trtP^\alpha
\partial^\beta\varphi \big]
\nn \\
{} & & + \frac{1}{4}\tF^2 \big[
(\trtP)^2 + (D-4)\trtP_\alpha\partial^\alpha\varphi \big] + \tP^2
\big[ (\trtP)^2 + (D-4)\trtP_\alpha \partial^\alpha\varphi \big]
\nn \\
{}Ê& & + \tr(\tP_\alpha\tP_\beta)\big[ \tr(\tP^\alpha\tP^\beta) -
\trtP^\alpha\trtP^\beta - 2(D-4)\trtP^\alpha
\partial^\beta\varphi \big]
\nn \\
{}Ê& & +
\frac{1}{2}(\trtP)^2\big[ (\trtP)^2 + 2(D-4)\trtP_\alpha
\partial^\alpha\varphi + (D-4)^2(\partial\varphi)^2 +
(D-4)\square\varphi \big] \big\}
\nn \\
{} & & + e\nabla_\alpha
\big\{ e^{(D-4)\varphi}\tvbint \big[ \big( - 2R^{\alpha\beta} +
\eta^{\alpha\beta}R +
\tF_{c\delta}^{\phantom{c\delta}\alpha}\tF^{c\delta\beta} +
2\tr(\tP^\alpha\tP^\beta)
\nn \\
{} & &
 - (D-4)\partial^\alpha\varphi\trtP^\beta \big)\trtP_\beta -
\big( \frac{1}{4}\tF^2 + \tr(D_\beta\tP^\beta) +
\tr(\tP_\beta\tP^\beta) + (\trtP)^2 \big)\trtP^\alpha \big]
\nn \\
{}Ê& & + \frac{1}{2}D^\alpha\big[ e^{(D-4)\varphi}\tvbint
(\trtP)^2 \big] \big\}.
\eqnlab{RicciSquared}
\eeqa
Using also $\tilde{\square}\varphi = \square\varphi +
\partial_\alpha\varphi\tr\tP^\alpha$ and $(\tilde{\partial}\varphi)^2
= (\partial\varphi)^2$ we have all the ingredients to extract the
compactified Gauss-Bonnet Lagrangian, \Eqnref{eqn:GBweyl2}. Notice
that $\hat{e}(\hat{\nabla}_A\hat{X}^A) =
\hat{\partial}_M(\hat{e}\hat{X}^M) =
\partial_\mu(\hat{e}\hat{X}^\mu)$ holds after the
compactification as well, implying that the total derivative terms
in \Eqnref{eqn:GBweyl2} will still be total derivatives even after
the compactification. Together with the result from the
Weyl-rescaling, \Eqnref{eqn:GBweyl2}, the complete result after the
compactification is
\beqa
{}Ê & & \hat{e}\hat{R}^{2}_{\mathrm{GB}} =
\sqrt{|g|}\tilde{e}_{\mathrm{int}}e^{(D-4)\varphi} \bigg\{ [
R_{\alpha\beta\gamma\delta}R^{\alpha\beta\gamma\delta} -
4R_{\alpha\beta}R^{\alpha\beta} + R^2 ] - \frac{1}{2}[
R_{\alpha\beta\gamma\delta} \tF_c^{\phantom{c}\alpha\beta}
\tF^{c\gamma\delta} + R\tF^2 \nn \\
{}Ê & & - 4R_{\alpha\beta}
\tF_{c\delta}^{\phantom{c\delta}\alpha} \tF^{c\delta\beta} ] +
\frac{1}{8}\tFFFFc -
\frac{1}{2}\tFFFFb + \frac{(D-n)}{16(D-n-2)}(\tF^2)^2 \nn \\
{}Ê & & + 2\tFPPFc + \tF^2 ( \frac{1}{2}(\trtP)^2 + (D-4)\trtP_\beta
\partial^\beta\varphi + \frac{(D-4)^2}{2}(\partial\varphi)^2 ) \nn
\\
{}Ê & & - \frac{1}{2} \tFPFa ( \trtP_\alpha +
(D-4)\partial_\alpha\varphi ) - 2\tFPFb ( \trtP_\alpha +
(D-4)\partial_\alpha\varphi ) \nn \\
{}Ê & & + 2\tr(\tP_\alpha\tP_\beta \tP^\alpha\tP^\beta) +
2\tr(\tP_\alpha\tP_\beta)\tr(\tP^\alpha\tP^\beta) - (\tP^2)^2 -
4(D-2)\tr(\tP_\alpha\tP_\beta)\partial^\alpha\varphi
\partial^\beta\varphi \nn \\
{}Ê & & - 4\tr(\tP_\alpha\tP^\alpha\tP^\beta) ( \trtP_\beta +
(D-4)\partial_\beta\varphi ) - 4(D-4)(\trtP)^2\trtP_\alpha
\partial^\alpha\varphi \nn \\
{}Ê & & + 2\tP^2 ( (\trtP)^2 + 2(D-4)\trtP_\alpha
\partial^\alpha\varphi + (D^2-7D+14)(\partial\varphi)^2) -
(\trtP)^2(\trtP)^2 \nn \\
{}Ê & & - 2(D^2-7D+14)(\trtP)^2(\partial\varphi)^2 - 4(D^2-8D+14)
(\trtP_\alpha \partial^\alpha\varphi)^2\nn \\
{}Ê & & - 4(D-4)(D^2-7D+11)\trtP_\alpha \partial^\alpha\varphi
(\partial\varphi)^2 - (D-2)(D-3)(D-4)(D-5)(\partial\varphi)^2
(\partial\varphi)^2 \nn \\
{}Ê & & + [R_{\alpha\beta} - \frac{1}{2}\eta_{\alpha\beta}R -
\frac{1}{2}\tF_{c\delta\alpha}\tF^{c\delta}_{\phantom{c\delta}\beta}
+ \frac{1}{8}\eta_{\alpha\beta}\tF^2 - \tr(\tP_\alpha \tP_\beta) +
\frac{1}{2}\eta_{\alpha\beta}\tP^2 + (D-2)\partial_\alpha\varphi
\partial_\beta\varphi
\nn \\
{}Ê & & - \frac{(D-2)}{2} \eta_{\alpha\beta} (\partial\varphi)^2 ] (
4\tr(\tP^\alpha \tP^\beta) - 4\trtP^\alpha \trtP^\beta -
8(D-4)\trtP^\alpha \partial^\beta\varphi
 \nn \\
{}Ê & & - 4(D-3)(D-4)\partial^\alpha\varphi
\partial^\beta\varphi ) \nn \\
{}Ê & & + [ \EOMtF ] (
-2\tF_d^{\phantom{d}\delta\gamma}\tP_\gamma^{\phantom{\gamma}cd} +
2\tF^{c\delta\gamma}\trtP_\gamma + 2(D-4)\tF^{c\delta\gamma}
\partial_\gamma\varphi ) \nn \\
{}Ê & & + [ \EOMtPa ]( - \frac{1}{2} \tF^c_{\phantom{c}\gamma\delta}
\tF^{d\gamma\delta} - 4\tP_{\gamma
e}^{\phantom{\gamma e}c} \tP^{\gamma ed}  \nn \\
{}Ê & & + 4\tP^{\gamma cd}\trtP_\gamma + 4(D-4) \tP^{\gamma
cd}\partial_\gamma\varphi ) \nn \\
{}Ê & & + \frac{1}{(D-2)} [ \EOMtrtP ] (
\frac{(D-3)}{2}\tF^2 + 2(D-4)\tP^2 - 2(D-4)(\trtP)^2  \nn \\
{}Ê & & - 4(D-3)(D-4)\trtP_\beta \partial^\beta\varphi -
2(D-2)(D-3)(D-4)(\partial\varphi)^2 ) \nn \\
{}Ê & & + 2(D-4)[ \EOMphi ] ( \frac{1}{4}\tF^2 +
\tP^2 - (\trtP)^2 - 2(D-3)\trtP_\beta \partial^\beta\varphi  \nn \\
{}Ê & &  - (D-2)(D-3)(\partial\varphi)^2 ) \bigg\} \nn \\
{}Ê & & + \mc{L}_{TD},
\eqnlab{GBtilde}
\eeqa
where the last term, $\mc{L}_{TD}$, is a total derivative which is
given explicitly by
\beqa
{}Ê& & \mc{L}_{TD} =  \sqrt{|g|} D^\alpha \bigg\{ D_\alpha \Big[
\tilde{e}_{\mathrm{int}}e^{(D-4)\varphi} ( \frac{1}{2}\tF^2 + 2\tP^2
- 2(\trtP)^2 )\Big]  +
\tilde{e}_{\mathrm{int}} e^{(D-4)\varphi} \Big[  \nn \\
{}Ê & & + 8[R_{\alpha\beta} - \frac{1}{2}\eta_{\alpha\beta}R -
\frac{1}{2}\tF_{c\delta\alpha}\tF^{c\delta}_{\phantom{c\delta}\beta}
+ \frac{1}{8}\eta_{\alpha\beta}\tF^2 - \tr(\tP_\alpha \tP_\beta) +
\frac{1}{2}\eta_{\alpha\beta}\tP^2 + (D-2)\partial_\alpha\varphi
\partial_\beta\varphi
\nn \\
{}Ê & & - \frac{(D-2)}{2} \eta_{\alpha\beta} (\partial\varphi)^2 ]
\trtP^\beta - 4[ D_\beta \tP^{\beta cd} - \frac{1}{4}
\tF^c_{\phantom{c}\beta\gamma}\tF^{d\beta\gamma} -
\frac{1}{(D-2)}\delta^{cd} \tr(D_\beta \tP^\beta) ] \tP_
{\alpha cd} \nn \\
{}Ê & & - 2 [ D_\gamma \tF^{c\delta\gamma} +
\tP_{\gamma}^{\phantom{\gamma}ce} \tF_{e}^{\phantom{e}\delta\gamma}
] \tF_{c\delta\alpha} + 4\frac{(D-3)}{(D-2)} [ \tr(D_\beta
\tP^\beta) - \frac{(D-2)}{4(D-n-2)}\tF^2 ]
\trtP_\alpha \nn \\
{}Ê & & + \frac{1}{2}\tF_{c\beta\gamma}\tP_\alpha^{\phantom{\alpha}
cd}\tF_d^{\phantom{d}\beta\gamma} + 2\tF_{c\beta\alpha}
\tP_\gamma^{\phantom{\gamma} cd}\tF_d^{\phantom{d}\beta\gamma} +
\frac{(n-1)}{(D-n-2)}\tF^2\trtP_\alpha -
(D-4)\tF^2\partial_\alpha\varphi \nn \\
{}Ê & & + 4\tr(\tP_\alpha\tP_\beta\tP^\beta) + 4( (\trtP)^2 -
\tr(\tP_\beta\tP^\beta) ) ( \trtP_\alpha +
(D-4)\partial_\alpha \varphi ) \nn \\
{}Ê & &   + (D^2 - 9D + 16) ( 4 \partial_\alpha \varphi
\trtP_\beta - 2 \trtP_\alpha \partial_\beta \varphi )
\partial^\beta \varphi \Big] \bigg\}.
\eeqa
All terms are thus grouped according to equations of motion and
total derivatives. The first two square parenthesis in
\Eqnref{GBtilde} -- containing terms involving only the Riemann
tensor and $\tF$ -- will vanish identically when compactifying to
three dimensions.

Varying the compactified Einstein-Hilbert action,
$\mathcal{L}_{\mathrm{EH}} = \hat{e}\hat{R}$, the tree-level
equations of motion are found to be
\beqa
{}Ê0 &=& R_{\alpha\beta} - \tP_{\alpha
cd}\tP_\beta^{\phantom{\beta}cd} - \frac{1}{2}\tF_{c\alpha\delta}
\tF^{c\phantom{\beta}\delta}_{\phantom{c}\beta} +
\frac{1}{4(D-n-2)}\eta_{\alpha\beta}\tF^2 +
(D-2)\partial_\alpha\varphi \partial_\beta\varphi,
\nn \\
{}Ê0 &=& D_\gamma\tF_a^{\phantom{a}\beta\gamma} + \tP_{\gamma
ad}\tF^{d\beta\gamma},
\nn \\
{}Ê0 &=& D_\gamma\tP^\gamma_{\phantom{\gamma}ab} -
\frac{1}{4}\tF_{a\gamma\delta}\tF_b^{\phantom{b}\gamma\delta} -
\frac{1}{4(D-n-2)}\delta_{ab}\tF^2.
\eqnlab{eom}
\eeqa
Notice that tracing the last equation in \Eqnref{eom}, one finds
$\square \varphi + \frac{1}{4(D-n-2)}\tF^2 = 0$ for the dilatons.
Except for the equations of motion, the fields will also obey the
Bianchi identities
\beq
\nabla_{[a}\tF^m_{\phantom{m}\beta\gamma]} = 0,
\eeq
and the Maurer-Cartan equations
\beqa
{}Ê0 &=& D_{[\alpha} \tP_{\beta] cd},
\nn \\
{}Ê0 &=& \nabla_{[\alpha} Q_{\beta] cd} -
Q_{[\alpha|c|}^{\phantom{\alpha |c|}e}Q_{\beta] de} +
\tP_{[\alpha|c|}^{\phantom{\alpha|c|}e}\tP_{\beta]de}.
\eeqa

\end{appendix}

%\end{thebibliography}

\end{document}